\documentclass[sn-mathphys-ay]{sn-jnl}

\usepackage{makecell}
\usepackage{graphicx}%
\usepackage{multirow}%
\usepackage{amsmath,amssymb,amsfonts}%
\usepackage{amsthm}%
\usepackage{subcaption}
\usepackage{siunitx}
\usepackage{mathrsfs}%
\usepackage[title]{appendix}%
\usepackage{xcolor}%
\usepackage{textcomp}%
\usepackage{manyfoot}%
\usepackage{booktabs}%
\usepackage{algorithm}%
\usepackage{algorithmicx}%
\usepackage{algpseudocode}%
\usepackage{listings}%

\theoremstyle{thmstyleone}%
\theoremstyle{thmstyletwo}%

\theoremstyle{thmstylethree}%

\newcommand\pbdot{\dot{P}_\mathrm{b}}{}

\begin{document}

\title[Unlocking Gravity and Gravitational Waves with Radio Pulsars: Advances and Challenges]{Unlocking Gravity and Gravitational Waves with Radio Pulsars: Advances and Challenges}


\author*[1]{\fnm{Huanchen} \sur{Hu}}\email{huhu@mpifr-bonn.mpg.de}

\affil*[1]{\orgdiv{Department of Fundamental Physics in Radio Astronomy}, \orgname{Max-Planck-Institut f\"ur Radioastronomie}, \orgaddress{\street{Auf dem H\"ugel 69}, \city{Bonn}, \postcode{53121}, 
\country{Germany}}}


\abstract{
Pulsars, the cosmic lighthouses, are strongly self-gravitating objects with core densities significantly exceeding nuclear density. 
Since the discovery of the Hulse--Taylor pulsar 50 years ago, binary pulsar studies have delivered numerous stringent tests of General Relativity (GR) in the strong-field regime as well as its radiative properties---gravitational waves (GWs). 
These systems also enable high-precision neutron star mass measurements, placing tight constraints on the behaviour of matter at extreme densities. 
In addition, pulsars act as natural detectors for nanohertz GWs, primarily from supermassive black hole binaries, culminating in the first reported evidence of a stochastic GW background in 2023. 
In this article, I review key milestones in pulsar research and highlight some of contributions from my own work. 
After a brief overview of the gravity experiments in \S\ref{intro}, I review the discovery of pulsars---particularly those in binaries---and their critical role in gravity experiments (\S\ref{sec2}) that laid the foundation for recent advances. 
In \S\ref{sec3}, I present the latest efforts on GR tests using the Double Pulsar and a pioneer technique to constrain the dense matter equation of state. 
\S\ref{sec4} demonstrates the potential of binary pulsars on testing alternative theories to GR. 
Advances in nanohertz GW detection with pulsar timing arrays are discussed in \S\ref{PTA}. 
I outline some of the current challenges in \S\ref{limit} and conclude with final remarks in \S\ref{conclude}. 
}


\maketitle

\section{Introduction to gravity experiments}\label{intro}

In 1915, Albert Einstein finalised his General theory of Relativity \citep[GR;][]{Einstein1915a}, revolutionised our view of gravity and spacetime. 
For many years, tests of GR were largely confined to the Solar System, a weak-field slow-motion regime of gravity.
Classical tests, such as the perihelion advance of Mercury \citep{Einstein1915b}, the deflection of starlight by the Sun \citep{Dyson+1920}, and gravitational redshift measurements \citep{adams1925,pound1959}, provided early confirmations of GR. 
In addition, Einstein's theory also predicted the existence of gravitational waves (GWs) generated by accelerated masses \citep{Einstein1916g,Einstein:1918btx}.
However, probing the radiative aspects of gravity and testing GR and its alternatives beyond the first post-Newtonian (PN) approximation remained unexplored until fifty years ago.

In the summer of 1974, Russell Hulse and Joseph Taylor discovered the first binary pulsar, PSR~B1913+16 \citep{HT1975ApJ}, which is at the same time the first double neutron star system. 
This system opened a completely new arena for studying gravity beyond the weak field of the Solar System, with a gravitational potential five orders of magnitude stronger. 
With this new laboratory and continued observations, for the first time ever, relativistic effects caused by the gravitational interaction of two strongly self-gravitating objects (i.e. neutron stars) can be studied.
One of the most profound early results from this system is that the orbital period decay is consistent with the rate of orbital energy loss due to GW emission as predicted by GR, providing the first evidence for the existence of GWs \citep{Taylor1982ApJ,Taylor1989ApJ}. 

The significance of this discovery extended far beyond binary pulsars. 
The test of GW emission from the Hulse--Taylor (HT) pulsar was a key milestone that ultimately led to the development of ground-based GW detectors such as LIGO and Virgo. 
These detectors made history in 2015 with the first direct detection of GWs from a binary black hole merger \citep{LIGO2016a}, opening an entirely new way of observing the Universe. 
Since then, the LIGO-Virgo-KAGRA (LVK) collaboration have detected numerous binary mergers, including double neutron star collisions, which provide complementary tests of GR in the highly dynamical strong-field regime \citep{LIGO2016GR,LIGO2019GR}. 
Furthermore, pulsars also serve as cosmic detectors for nanohertz GWs, primarily from supermassive black hole binaries (SMBHBs), the loudest GWs in the Universe. 
The pulsar timing array (PTA) experiments are now approaching the long-anticipated detection of ultra-low-frequency GWs, further broadening the scope of GW astrophysics \citep{EPTA2023A&AIII, CPTA2023RAA, NanoGrav2023ApJ, PPTA2023ApJ, MPTA2025}.

Beyond GWs, the Event Horizon Telescope (EHT) has provided the first direct images of black hole shadows, offering new tests of GR near event horizons. 
Observations of M87* and Sagittarius A* have placed constraints on deviations from the Kerr metric predicted by GR \citep{EHT2019_M87,EHT2022_SgrA}. 
Meanwhile, various Solar System tests have been significantly improved in precision. 
Very-long-baseline interferometry (VLBI) observations \citep{Shapiro2004_GRTests,Fomalont+2009}, have confirmed gravitational deflection in agreement with GR at a precision of $1.5\times 10^{-4}$. 
Measurements from the Cassini spacecraft verified the Shapiro delay \citep{Shapiro1964} to an accuracy of $10^{-5}$ with GR \citep{Bertotti2003_Cassini}. 
Additionally, GR has successfully passed numerous other tests, including the Lunar Laser Ranging experiment for the de Sitter precession of the Moon's orbit and the strong equivalence principle \citep{Nordtvedt1999}, the Gravity Probe B for the geodetic and frame-dragging effects \citep{GPB2011}, and the Lense--Thirring effect with satellite laser ranging \citep{Ciufolini2004,Ciufolini_2019}.

Despite these achievements, pulsars play a crucial---and sometimes unique---role in gravity tests. 
They are powerful tools for addressing two fundamental questions: Is GR a complete description of gravity on macroscopic scales? And how does matter behave at extreme densities?
Radio observations of binary and triple systems have placed some of the most stringent limits on possible deviations from GR, as well as provided precise neutron star mass measurements that have helped to constrain the equation of state of dense matter \citep{Wex2020Univ,Freire2024LRR}. 
These all rest on the foundations established by early discoveries---from Jocelyn Bell Burnell's identification of the first pulsar to the groundbreaking detection of the HT pulsar.
In the following section, I will review these pioneering efforts that paved the way for today's pulsar experiments in understanding gravity and GWs.

\section{58 years of pulsars and 50 years of binary pulsars}\label{sec2}

\begin{figure}[t]
    \vspace{-10pt}
    \centering
    \begin{subfigure}{0.49\textwidth}
    \includegraphics[width=\textwidth]{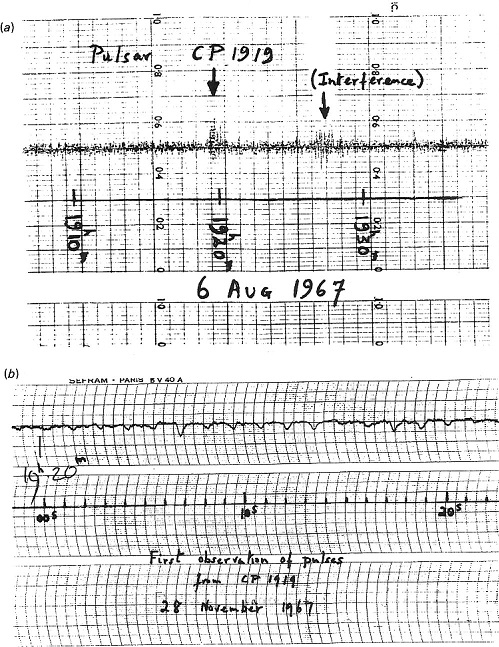}
    \caption{}
    \label{fig:chart}
    \end{subfigure}
    \begin{subfigure}{0.5\textwidth}
    \includegraphics[width=\textwidth]{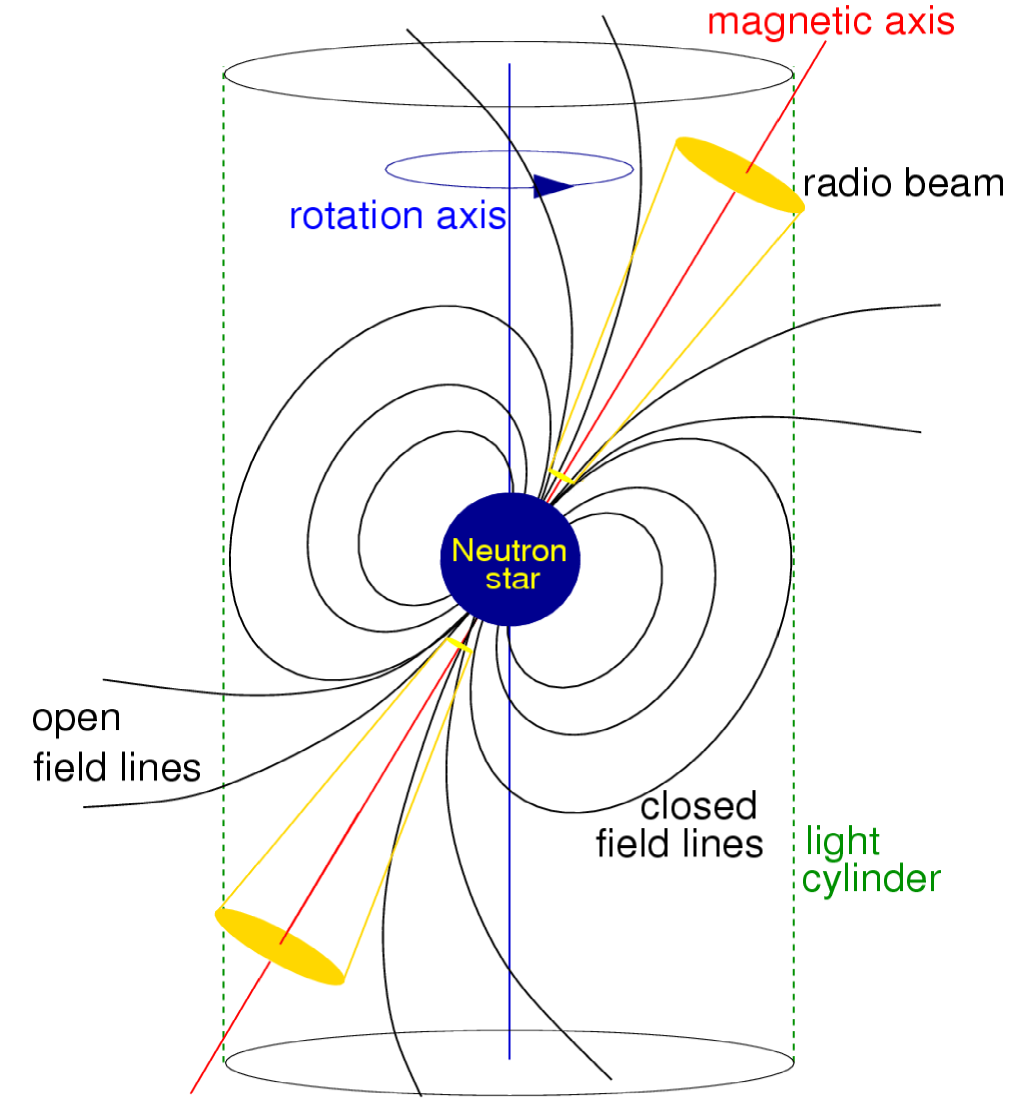}\vspace{20pt}
    \caption{}
    \label{fig:psr}
    \end{subfigure}
    \caption{(a) Two pen chart recordings taken by Jocelyn Bell Burnell during her PhD \citep{Hewish1968}. It shows the first detection of a pulsar on 6 August 1967, which was subsequently confirmed on 28 November 1967 with high-speed recordings. Credit: CSIROpedia. (b) A simplified pulsar model adapted from \citet{LK2004}.}
    \label{fig:pulsar}
\end{figure}

The discovery of pulsars dates back to 1967, when Jocelyn Bell Burnell, while working on a PhD project on quasar scintillation with Anthony Hewish, observed repeating radio pulses using a pen chart recorder on the antennas she had constructed at the Mullard Radio Astronomy Observatory outside Cambridge (see Fig.~\ref{fig:chart}). 
The timescales of such repetitive signals are much shorter than those expected for quasar scintillation, which are the result of the interaction of radio waves with the ionised interstellar medium on their way.
Subsequent measurements confirmed that the signals originated from a fixed declination and right ascension, indicating an extraterrestrial source. 
High-speed chart recordings in Fig.~\ref{fig:chart} revealed that the pulses recurred at precise 1.337-s intervals with remarkable stability \citep{Hewish1968}.
Soon enough, more similar sources was identified by Bell Burnell from her chart recordings, indicating that they belonged to a new class of astronomical objects, commonly known as ``pulsars'' today. 

Around the same period, Franco Pacini and Thomas Gold independently proposed that pulsed radiation can be originated from rotating magnetised neutron stars produced during supernova explosions of massive stars \citep{Pacini1967,Gold1968}. 
When a massive star ($\gtrsim 10~\mathrm{M_\odot}$) exhausts its nuclear fuel at the end of its life, its core collapses under its own gravity. Protons and electrons merge into neutrons, resulting in a super-dense neutron-rich core supported by neutron degeneracy pressure, called a ``neutron star'', with a mass greater than $1.4~\mathrm{M_\odot}$ within a radius of $\sim$12~km \citep{Chandrasekhar1931,Landau1932}. 
This collapse triggers a supernova explosion in which the outer layers rebound from the inner core and are ejected in a powerful explosion that enriches space with heavy elements \citep{BZ1934a,BZ1934c}. 
The first human record of supernova explosion was made by Chinese astronomers in 1054 AD, which created the Crab Nebula.
Shortly after Bell Burnell's discoveries, a 33-ms pulsar was found at the centre of the Crab Nebula \citep{SR1968}, further supporting the rotating neutron star model of pulsars (see Fig.~\ref{fig:psr} for an illustration). 
As it rotates, the emission beam along the magnetic axis sweeps across the Earth and leaves regularly spaced pulses, which could be used as clocks. 

By monitoring the times of arrival (TOAs) of radio pulses, a phase-connected timing solution can be used to account for every rotation of the pulsar and allow high precision measurements of timing parameters that encode information about pulsar itself (e.g. spin and astrometry) and the interstellar medium along the way. This process is known as ``pulsar timing'' technique \citep[for a more complete overview see e.g.][]{LK2004}.

Being the most compact objects after black holes, combined with its stable periodicity, these strongly self-gravitating bodies are ideal probes for a wide range of questions in fundamental physics, in particular if the pulsar is in a binary system. 
The first of such systems was discovered fifty years ago by Russell Hulse and Joseph Taylor in a pulsar survey carried out at the Arecibo Observatory. 
Rapid periodic changes in its spin period revealed that the pulsar is in a binary system \citep{HT1975ApJ}.
The HT pulsar B1913+16 was soon recognized as an excellent laboratory for testing gravity, opening a new era for studying relativistic effects in both the quasi-stationary strong-field regime and the radiation regime \citep{Wex_2014}.

\begin{figure}[t]
    \centering
    \begin{subfigure}{0.48\textwidth}
    \includegraphics[width=\textwidth]{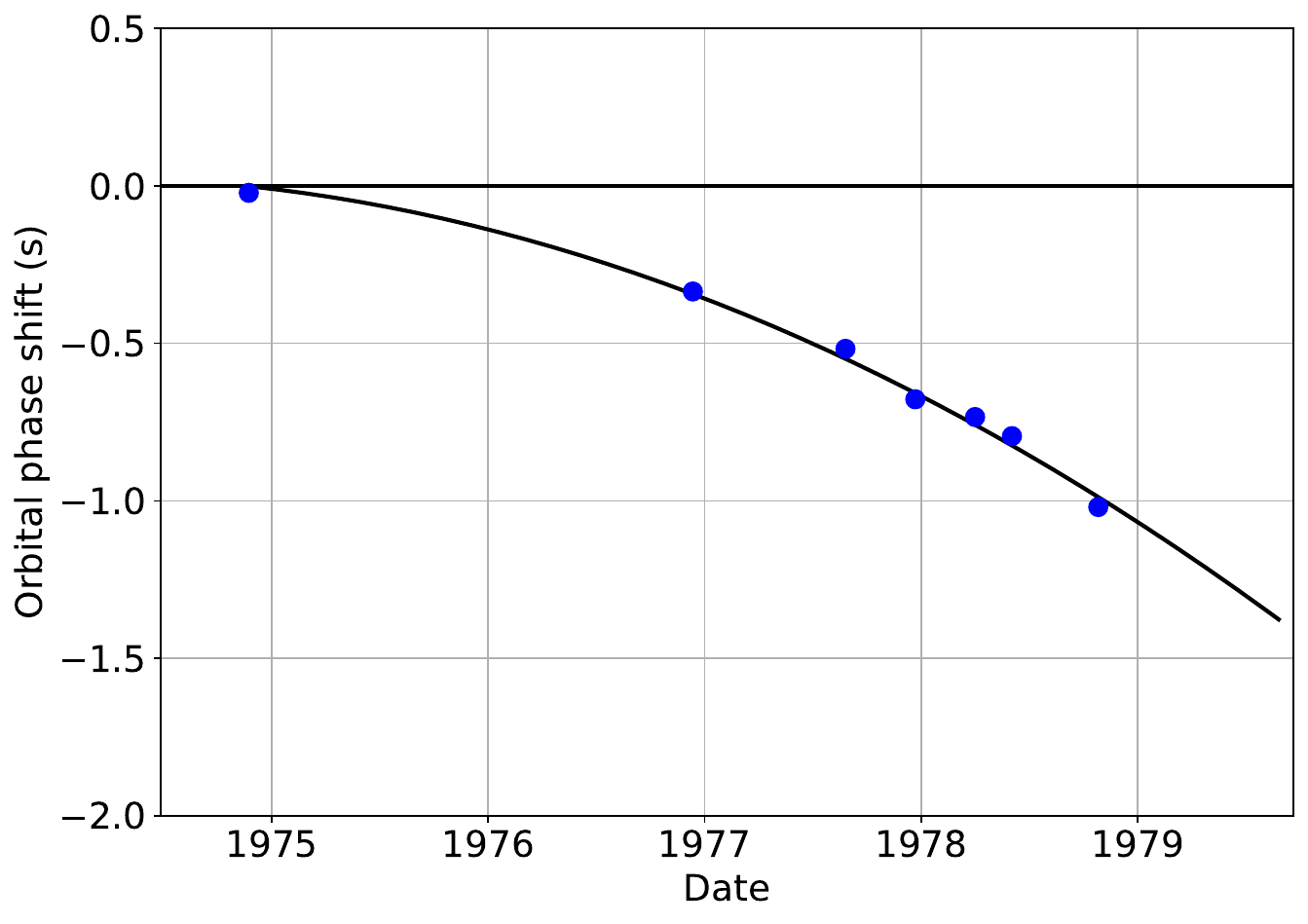}\vspace{60pt}
    \caption{}
    \label{fig:HT}
    \end{subfigure}
    \begin{subfigure}{0.5\textwidth}
    \includegraphics[width=\textwidth]{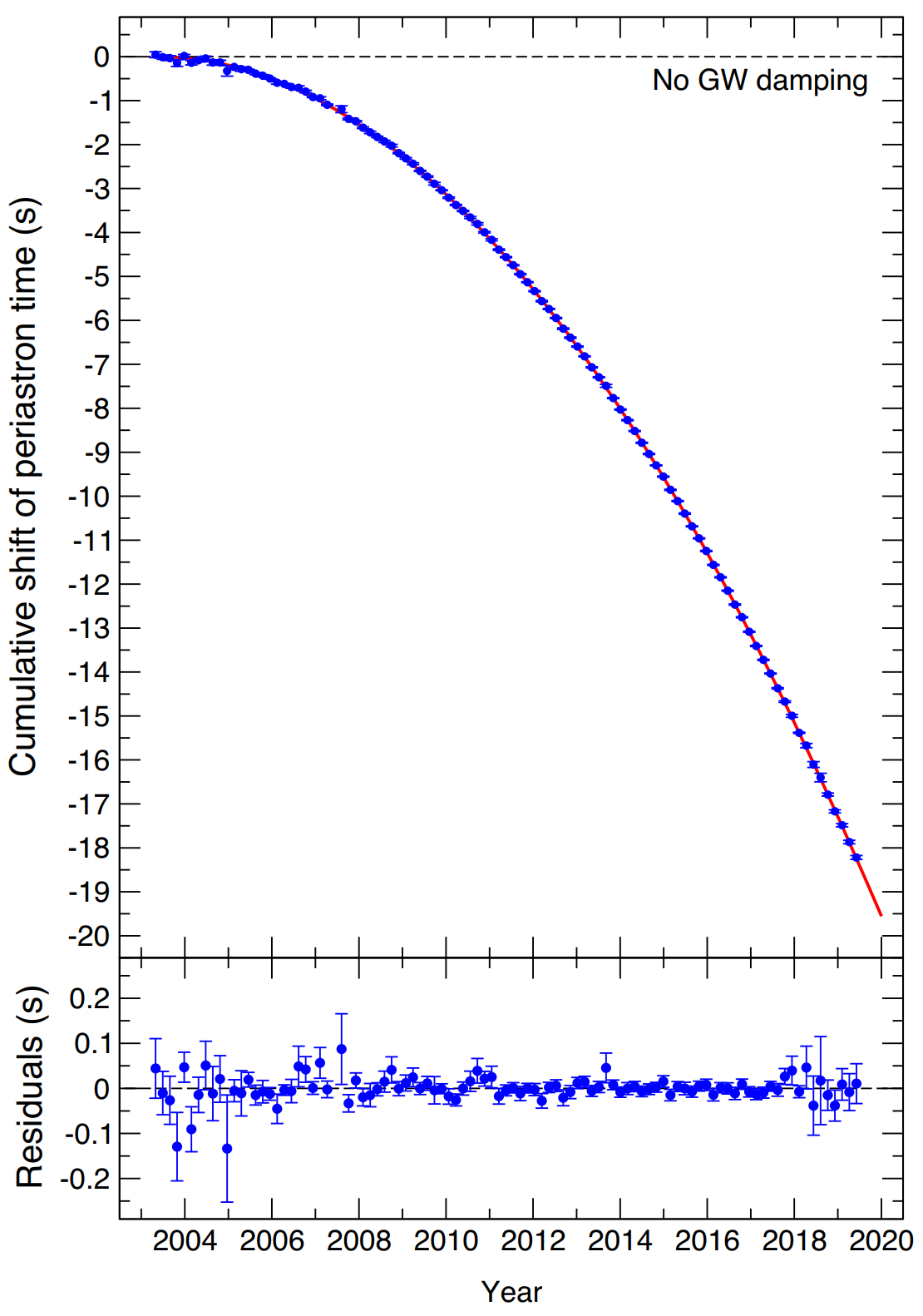}
    \caption{}
    \label{fig:DP}
    \end{subfigure}
    \caption{Orbital period decay in PSR~B1913+16 (left) and PSR~J0737$-$3039A/B (right) demonstrated as an increasing orbital phase shift for periastron passages with time. The parabolas indicate GR's prediction due to GW emission. (a) First evidence of the existence of GWs \citep{Taylor1979Natur}. Figure courtesy of Norbert Wex, replotted using WebPlotDigitizer \citep{WebPlotDigitizer}. (b) The bottom shows the deviation from GR (red). Reproduced from \cite{Kramer+2021PRX} under CC BY 4.0.}
    \label{fig:gw}
    \vspace{-10pt}
\end{figure}

Following this discovery, timing models were constructed to parameterise Keplerian orbit of binary pulsars and relativistic deviations, the later of which are commonly referred to as post-Keplerian (PK) parameters \citep{BT1976, DD85, DD86}. 
For a given gravity theory, such as GR, PK parameters are functions of Keplerian parameters and a priori unknown pulsar mass $m_\mathrm{p}$ and companion mass $m_\mathrm{c}$. 
Three PK parameters were soon measured in the HT pulsar: the advance of periastron $\dot{\omega}$, Einstein delay $\gamma_\mathrm{E}$, and orbital period decay $\dot{P}_\mathrm{b}$ \citep{Taylor1979Natur}. 
With the first two, the masses of the pulsar and companion were determined to be $1.44\,\mathrm{M}_\odot$ and $1.39\,\mathrm{M}_\odot$ under the framework of GR, indicating that the companion is also a neutron star. 
Basing on these masses, GR's prediction of energy loss due to GW emission (at the quadruple order) was found to be consistent with observations as presented in Fig.~\ref{fig:HT}, providing the first evidence of the existence for GWs.
Further observations of the system revealed that the observed value of the orbital period decay rate $\dot{P}_\mathrm{b}$ agrees with GR's prediction at the level of about 1\% \citep{Taylor1982ApJ, Taylor1989ApJ}. 
After 35 years of timing, the fractional uncertainty of the GW test made with the HT pulsar has been lowered to $10^{-3}$ level \citep{WH2016ApJ,Deller+2018ApJ}.
Since then, many binary pulsars have been discovered, playing a crucial role in testing gravity and providing unique opportunities for the precise measurement of neutron star masses---and, potentially, moments of inertia---which help to constrain the equation of state (EOS) of dense matter \citep[see][for recent reviews]{Freire2024LRR, Hu2024Univ}.

Another milestone in pulsar studies is the discovery of the first millisecond pulsar (MSP), PSR~B1937+21, with a rotational period of only 1.558~ms \citep{Backer+1982Natur}. 
It is worth noting that MSPs are typically formed during a mass transfer from the companion star to the pulsar, spinning up the pulsars from a few seconds to a few milliseconds. 
Pulsars that underwent this process are called ``recycled'' pulsars, which have much smaller spin down rates ($\dot{P}/P\sim 10^{-20}\, \mathrm{s\,s^{-1}}$) than normal pulsars ($\dot{P}/P\sim 10^{-15}\, \mathrm{s\,s^{-1}}$) and thereby act as stable clocks.
A few years before the discovery, \cite{Sazhin1978} and \cite{Detweiler1979} already proposed the idea of using pulsars to search for GWs from supermassive binaries with periods of 1--10 years. 
As GWs propagate, they distort spacetime, causing significant advances or delays in the TOAs over long timescales, manifesting as a temporally correlated (red) signal.
Given their remarkable rotational stability, MSPs are ideal tools for this experiment. 

However, pulsar's intrinsic spin noise also acts as a red signal and causes difficulty in disentangling GWs. 
To overcome this problem, \cite{FB1990ApJ} first purposed to construct a spatial array of MSPs called a ``pulsar timing array (PTA)'' and search for a common red signal in all pulsars. 
The GWs passing over the Earth would leave a spatially correlated signal between pulsar pairs described by the Hellings--Downs (HD) correlation \citep{HD1983}. 
The recent advances on this topic is summarised in Section~\ref{PTA}.

\section{Double Pulsar's gravity dance}\label{sec3}

A gravity laboratory even better than the HT pulsar is the so-called Double Pulsar system, PSR~J0737$-$3039A/B, owing to the fact that it is the only system known to host two pulsars. 
The system is composed of a 23-ms ``recycled'' pulsar A and a 2.8-s ``young'' pulsar B in a mildly eccentric ($e=0.088$), edge-on, and tight orbit with a period of 2.45~h \citep{Burgay+2003,Lyne+2004}. 
These features make it a unique gravitational laboratory in many aspects \citep{Kramer+2006Sci}.

\begin{figure}[t]
    \centering
    \includegraphics[width=0.9\textwidth]{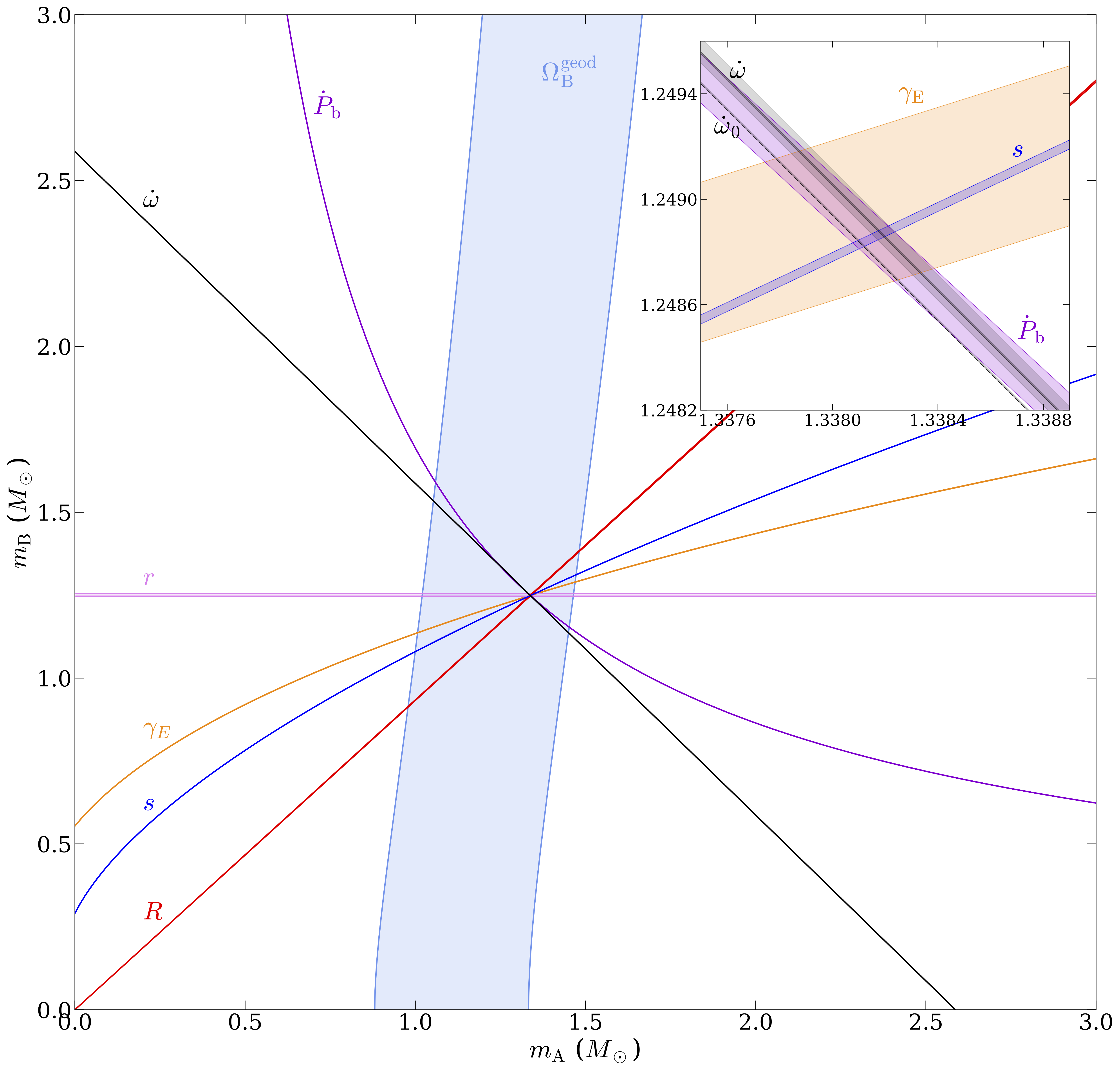}
    \vspace{5pt}
    \caption{Mass-mass diagram of the Double Pulsar plotted for six PK parameters and the mass ratio $R$ based on the assumption of GR. Each pair of curves shows $\pm 1 \sigma$. An enlarged view is shown in the upper right (only four parameters with the smallest uncertainties are plotted), where all the parameters are intersected with each other. Reproduced from \cite{Kramer+2021PRX} under CC BY 4.0.}
    \label{fig:mmdiag} 
\end{figure}

The ratio of the semi-major axes of two pulsars, $x_\mathrm{B}$ and $x_\mathrm{A}$, provided a unique mass ratio parameter $R \equiv m_\mathrm{A}/m_\mathrm{B} = x_\mathrm{A}/x_\mathrm{B}$ \citep{Perera+2010ApJ}, which is theory independent at the first order.
Due to relativistic spin precession (or ``geodetic precession'' $\Omega_\mathrm{B}^\mathrm{geod}$), pulsar B moved out of the view since 2008 and is predicted to be back around 2035 \citep{Breton+2008,Lower+2024A&A}.
Thanks to its much more compact orbit, relativistic effects in the orbital dynamics ($\dot{\omega}, \gamma_\mathrm{E},\dot{P}_\mathrm{b}$) are much more significant compared to the HT pulsar.
In addition, because of the nearly edge-on orbit viewed from the Earth (inclination angle $i \sim 90\si{\degree}$), when pulsar A's signal passes in the vicinity of pulsar B, it suffers a prominent delay by the gravitational field of B, known as the Shapiro delay \citep{Shapiro1964}. The measured Shapiro shape parameter ``$s$'' and range parameter ``$r$'' provide additional tests of GR. 
Using 16 years of data from six telescopes, the five classical PK parameters ($\dot{\omega}, \gamma_\mathrm{E}, \dot{P}_\mathrm{b}, r, s$) have been measured to a very high precision, and a new effect--- relativistic deformation of the orbit $\delta_\theta$---was also detected \citep{Kramer+2021PRX}. 
Along with $R$ and $\Omega_\mathrm{B}^\mathrm{geod}$, these parameters (except for $\delta_\theta$) are plotted as relations between two masses ($m_\mathrm{A}$ and $m_\mathrm{B}$) under the framework of GR in the so-called ``mass-mass diagram'' (see Fig.~\ref{fig:mmdiag}). 
All these parameters intersect perfectly, in other words, GR has passed all these tests. 

In particular, as presented in Fig.~\ref{fig:DP}, with the (GR) masses measured using $s$ and $\dot{\omega}$, the GW damping $\dot{P}_\mathrm{b}^\mathrm{GW}$ is in agreement with the Double Pulsar at a level of $1.3 \times 10^{-4}$ (95\% confidence). This yielded by far the most precise test of GR's quadrupolar description of GWs---approximately 25 times better than in the HT pulsar \citep{WH2016ApJ} and three orders of magnitude better than from GW merger events detected by ground-based GW detectors \citep{LIGO+V2019,LIGO+V2021}.

\subsection{State-of-the-art GR tests}
\label{sec31}

Despite the success that were already made with 16 years of the Double Pulsar data from six telescopes, the boundary of GR still remains untouched. 
Only with greatly improved sensitivity can the time to achieve a breakthrough be greatly reduced. 
This was exactly why the Square Kilometre Array (SKA) and its mid-frequency precursor, the MeerKAT telescope, play an important role here. 
The MeerKAT telescope, located in South Africa, consists of 64 antennas, each with a diameter of 13.5~m, which is equivalent to a 108-m telescope. 
It started operation in 2019 and regular observations of PSR~J0737$-$3039A were conducted as part of the Relativistic Binaries (``Relbin'') working group \citep{Kramer+2021Relbin} of the MeerTime collaboration \citep{Bailes2020}. 
Given that PSR~J0737$-$3039A is located in the Southern sky, MeerKAT and the SKA have a better geographic advantage than other large radio telescopes located mainly in the northern hemisphere. 
By analysing early MeerKAT observations, it already exhibited a great precision---approximately seven times better than the 64-m Parkes ``Murriyang'' telescope in Australia and three times better than the 100-m Green Bank Telescope (GBT) in West Virginia \citep{Hu+2022}. 

With this unparalleled precision, the measurements of the Shapiro delay parameters were much improved with only three years of MeerKAT data compared to those based on the 16-yr data. More specifically, the shape parameter $s$ was improved by two times \citep{Hu+2022} and the range parameter $r$ validated GR at a level of $5.3 \times 10^{-3}$ with 95\% confidence. 
Although the precision of long-term PK parameters measured from 3-yr MeerKAT data such as the advance of periastron $\dot{\omega}$ has not yet caught up with the 16-yr data, the combination of $s$ and $\dot{\omega}$ already yielded improved mass measurements of both pulsars: $m_\mathrm{A}=1.338186\pm 0.000010\,\mathrm{M}_\odot, m_\mathrm{B}=1.248866\pm 0.000007\, \mathrm{M}_\odot$.\footnote{Errors are shown at $1\sigma$ level.}
These are by far the most precise mass measurements of neutron stars. 

Moreover, next-to-leading order (NLO) signal propagation effects were also independently tested with the MeerKAT data. 
These include: 1) the retardation effect caused by the movement of pulsar B while A's signal propagates through the binary system, also known as the 1.5 PN correction to the Shapiro delay \citep{KS1999,RL2006_lensing}; 2) the deflection of A's signal by the gravitational field of B, formally referred to as the lensing correction to the aberration delay \citep{DK1995,RL2006_rotation}; and 
3) the lensing correction to the Shapiro delay \citep{KZ2010}, as the leading-order expression were based on integration along a straight path \citep{BT1976}. 
The combination of these three effects are fitted using a scaling factor $q_\mathrm{NLO}$ in the modified \textsc{DDS} timing model after \cite{DD86}.\footnote{A logarithmic Shapiro shape parameter $z_s \equiv -\ln(1-s)$ is used instead of $s$. See \cite{Kramer+2021PRX} for details on this timing model.}

\begin{figure}[t]
    \vspace{-15pt}
    \centering
    \includegraphics[width=0.8\textwidth]{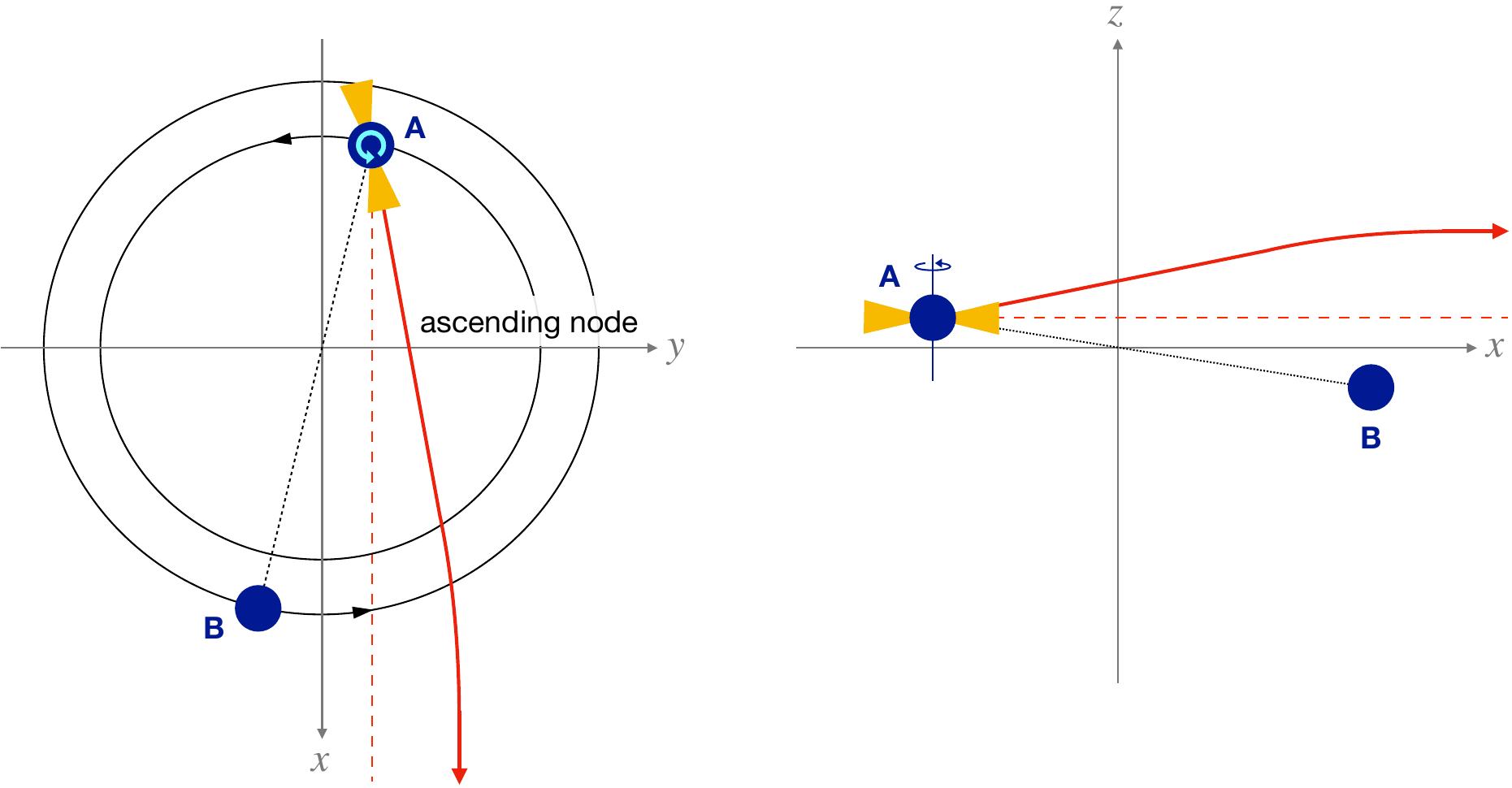}
    \vspace{5pt}
    \caption{Illustration of deflection effects of pulsar A's signal by the gravitational field of pulsar B both in longitude (left) and latitude (right). Reproduced from \cite{Hu+2022} under CC BY 4.0.}
    \label{fig:deflection}
    \vspace{-10pt}
\end{figure}

\begin{figure}[t]
    \vspace{-10pt}
    \centering
    \begin{subfigure}{0.49\textwidth}
    \includegraphics[width=\textwidth]{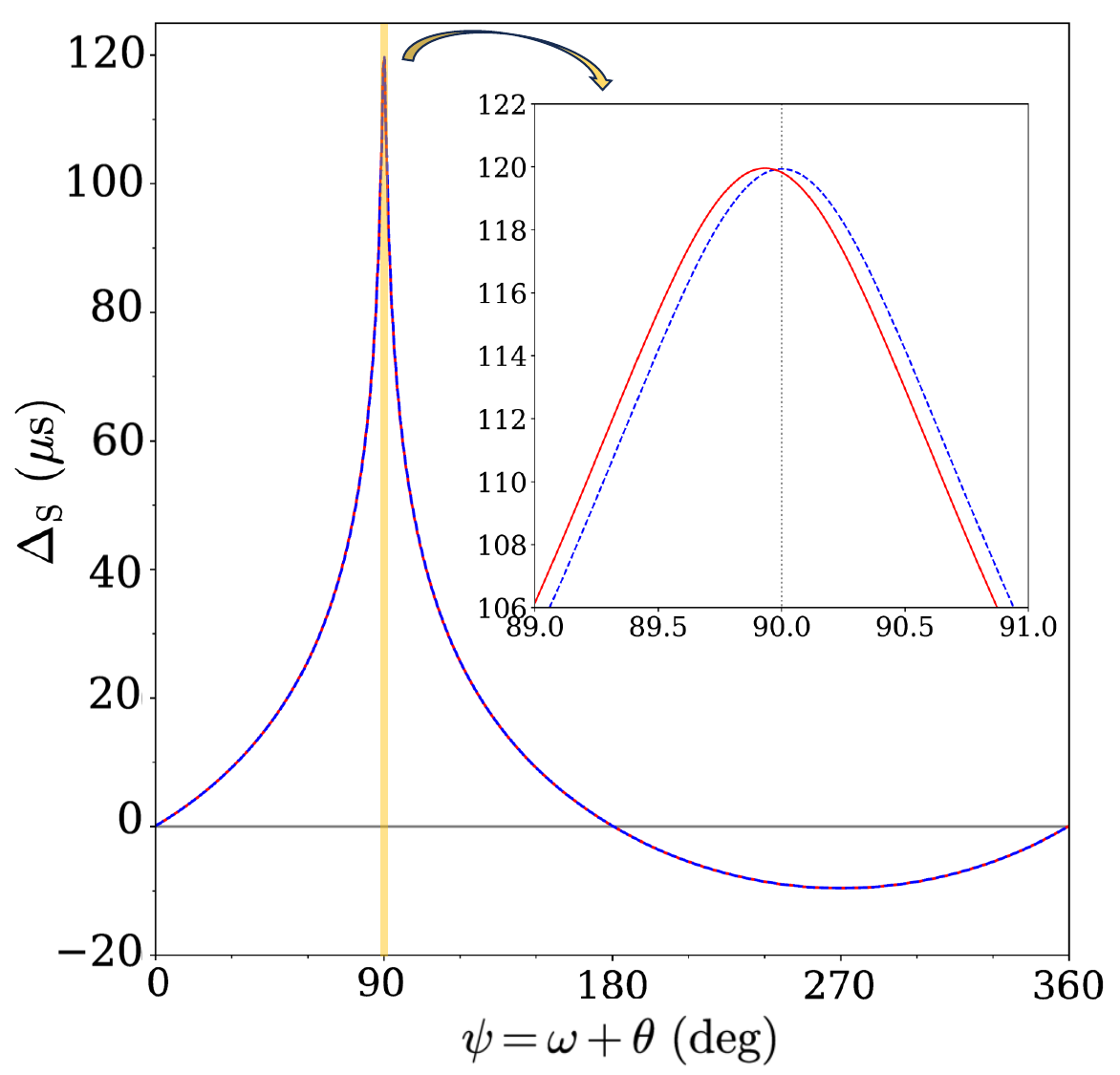}
    \caption{Shapiro delay with 1.5 PN correction.}
    \label{fig:shapiro}
    \end{subfigure}
    \begin{subfigure}{0.5\textwidth}
    \includegraphics[width=\textwidth]{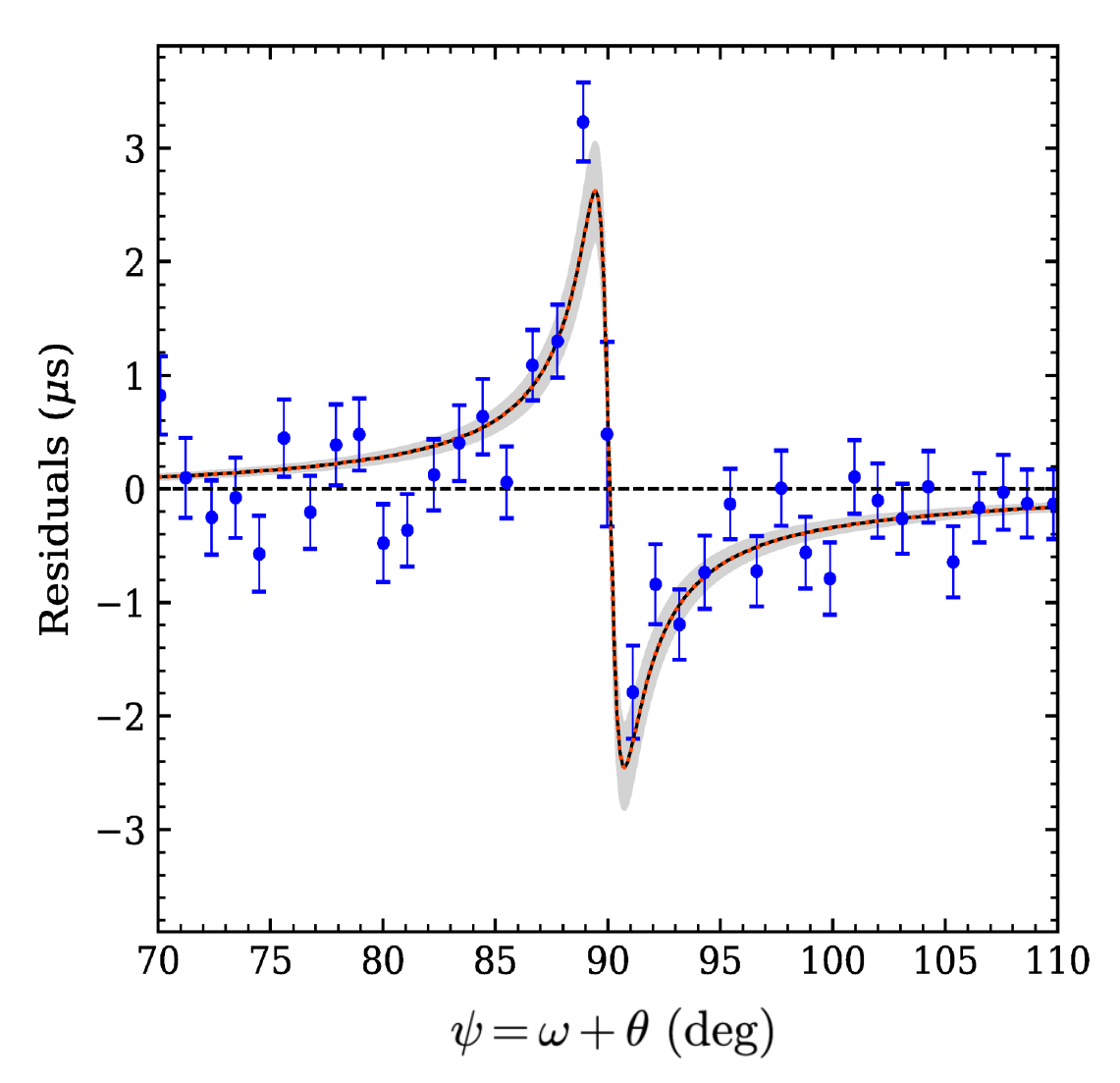}
    \caption{NLO signal propagation effects.}
    \label{fig:shaphof}
    \end{subfigure}
    \caption{Shapiro delay and NLO signal propagation effects. (a) 
    The Shapiro delay is shown as the dashed blue line, which is centred at the orbital phase of 90$\si\degree$. The retardation effect (1.5 PN) causes the curve to shift slightly forward, shown in red. Courtesy of Norbert Wex.
    (b) Added residuals (blue) due to NLO contributions in Shapiro and aberration delay with respect to orbital phase $\psi$. The black curve indicates the fitted value of $q_\mathrm{NLO}$ with $2\sigma$ uncertainty shown by the grey areas, which is in perfect agreement with the theoretical prediction of GR in red. Adapted from \cite{Hu+2022} under CC BY 4.0.}
    \label{fig:shap}
    \vspace{-10pt}
\end{figure}

An illustration of the geometry is shown in Fig.~\ref{fig:deflection}, where the spin axis of pulsar A is almost perpendicular to the orbital plane.
For the retardation effect, pulsar B acts as a moving lens, effectively advancing the Shapiro delay from the blue dashed curve in Fig.~\ref{fig:shapiro} to the red curve. 
If this is not taken into account, the difference between the two curves results in antisymmetric residuals as shown in Fig.~\ref{fig:shaphof}. The (longitudinal) deflection effect would result in a similar manner. 
This effect takes place because of the ``lighthouse'' behaviour of pulsars and acts as a NLO contribution to aberration delay \citep[see e.g.][for details]{Kramer+2021PRX}. 
On the left of Fig.~\ref{fig:deflection}, the rotation of pulsar A is prograde with its orbital motion (both counter-clockwise). 
Because of gravitational deflection, A's signal is bent by B's gravitational field, following the red solid path instead of the straight dashed path.
Therefore, the signal we receive before the superior conjunction (when A and B are aligned to line of sight) is delayed because A has to rotate an extra angle. 
Similarly, after the superior conjunction, A turns at a smaller angle and the signal is advanced. 
This is also consistent with the signature shown in the residual plot (Fig.~\ref{fig:shaphof}). 
Both the retardation and the longitudinal deflection effects have similar magnitude and constitute the pattern seen in the residual plot. 
This independently confirms the prograde rotation of pulsar A---if A's rotation were reversed, the longitudinal deflection would have the opposite effect, cancelling out the retardation effect. 
On the other hand, the lensing correction to the Shapiro delay is expected to show an advanced pattern symmetric to the superior conjunction ($\psi=90\si\degree$) as the curved path results in a reduced propagation time. 
However, its magnitude is much smaller ($\sim 0.8\mu$s) and it is highly correlated with the Shapiro shape parameter $s$, therefore remain undetected in any pulsar systems. 
Nevertheless, for completeness this contribution is included in the fitting of $q_\mathrm{NLO}$.
The 3-yr MeerKAT data yielded a test of these NLO signal propagation effects with $q_\mathrm{NLO}=0.999(79)$,\footnote{Errors shown in the brackets correspond to $1\sigma$, same for the rest of the paper.} which confirms GR ($q_\mathrm{NLO}=1$) with an accuracy of 16\% with 95\% confidence \citep{Hu+2022}, 1.6 times better than that of the 16-yr data \citep{Kramer+2021PRX}. 
These NLO effects can currently only be tested in the Double Pulsar system. 

When pulsar A is eclipsed by pulsar B, A's signal passes B's gravitational field as close as $\sim 10000$~km, where the spacetime curvature probed by photon propagation is $10^3$ and $10^9$ times larger than those probed
by black hole imaging around M87 \citep{EHT2019_M87} and Saggitarius A* \citep{EHT2022_SgrA}, and about $10^6$ times larger than similar experiments performed in the Solar System \citep{Bertotti2003_Cassini}. 
In fact, among all current gravity experiments that test photon propagation, the Double Pulsar probes the strongest spacetime curvature ($\sim 10^{-21}\,\mathrm{cm}^{-2}$).

In addition to the deflection at longitudinal direction, the lensing correction to the aberration delay can also cause a change in the co-latitude of the emission direction towards the Earth. 
An illustration is shown on the right of Fig.~\ref{fig:deflection}, and it is expected to cause profile variation, as a different region of pulsar beam enters the line of sight \citep{RL2006_lensing,RL2006_rotation}, and consequently a latitudinal deflection delay \citep{DK1995,RL2006_lensing}.
Evidence of profile variation at the superior conjunction has not been found in 3-yr MeerKAT data \citep{Hu+2022}, but may be possible with more data collected using MeerKAT and the SKA.

These results represent a notable test case of the capability of the MeerKAT telescope as a precursor for SKA science, and the ongoing data combination of 6-yr MeerKAT data with 16-yr data promises to take gravity tests to a new level of precision and enables the first measurement of neutron star moment of inertia (Hu et al. in prep.).

\subsection{Gravity and matter in the SKA era}
\label{sec32}

To predict the capability of the SKA in improving these gravity tests, I make simulations for MeerKAT, MeerKAT+, and the SKA as an update to the one made in 2020 \citep{Hu+2020}, as the plan regarding the construction and operation of MeerKAT+ and the SKA has been changed considerably. 
This simulation is performed in accordance with the upcoming new version of the SKAO science book.\footnote{\url{https://www.skao.int/en/science-users/557/advancing-astrophysics-ii}}

MeerKAT+ is an expanded phase of MeerKAT with the addition of 20 SKA-sized dishes, each 15~m in diameter. 
It is expected to be ready at the beginning of 2026. 
From the middle of 2028, the SKA Array Assembly phase AA* is expected to be finished, consisting of 64 MeerKAT dishes and 80 SKA dished.\footnote{\url{https://www.skao.int/en/science-users/599/scientific-timeline}} 
The full design baseline of the SKA phase 1 is called AA4, which has 53 additional SKA dished compared to AA*, and its timeline has not be determined. 
For the simulation, I simulate MeerKAT data from 2019/Q2 to 2025/Q4, MeerKAT+ data from 2026/Q1 to 2028/Q2, and both AA* and AA4 data from 2028/Q3 for a period of ten years as a comparison. 
The detailed simulation setup is summarised in Table~\ref{tab:sim}, based on the averaged timing precision and observing cadence from MeerKAT observations at two observing bands \citep{Hu+2022}.

\begin{table}[t]
{\renewcommand{\arraystretch}{1.5}
\centering
\caption{\small Summary of simulation setup for PSR~J0737$-$3039A based on MeerKAT observations. The considered observing frequency bands include the UHF band (544–1088~MHz), L band (856–1712~MHz), Band 1 (350–1050~MHz), and Band 2 (950–1760~MHz). The TOA uncertainty are estimated from MeerKAT observations and scaled to a sub-integration of 30~s over the full bandwidth. 
}\label{tab:sim}
\begin{tabular}{@{}lcccc@{}}
\toprule
Telescope & MeerKAT & MeerKAT+ & SKA-Mid AA* & SKA-Mid AA4 \\ 
\midrule
Number of dishes & 64 & 64/84 & 144 & 197 \\ 
Effective diameter [m] & 108  & 111 & 172 & 204 \\ 
Observing period [yr] & \makecell{2019/Q2-\\2025/Q4}  & \makecell{2026/Q1-\\2028/Q2} & \makecell{2028/Q3-\\2038/Q2} & \makecell{2028/Q3-\\2038/Q2} \\ 
Observing bands & UHF$|$L & UHF$|$L & 1$|$2 & 1$|$2 \\ 
TOA uncertainty [$\mu\rm{s}$] & 1.6$|$2.8 &  1.5$|$2.7 & 0.6$|$1.1 & 0.5$|$0.8  \\ 
\botrule
\end{tabular}
\footnotetext{
Note: MeerKAT+ enhances the existing MeerKAT array by adding 20 SKA dishes (15~m each) to the 64 MeerKAT dishes (13.5~m each), bringing the total to 84. However, the current beamforming pipeline supports only 64 dishes. Therefore, in MeerKAT+ simulations, 20 MeerKAT dishes are replaced with 20 SKA dishes to deliver a slightly better performance.}
}
\end{table}

\begin{figure}[t]
    \centering
    \includegraphics[width=0.75\textwidth]{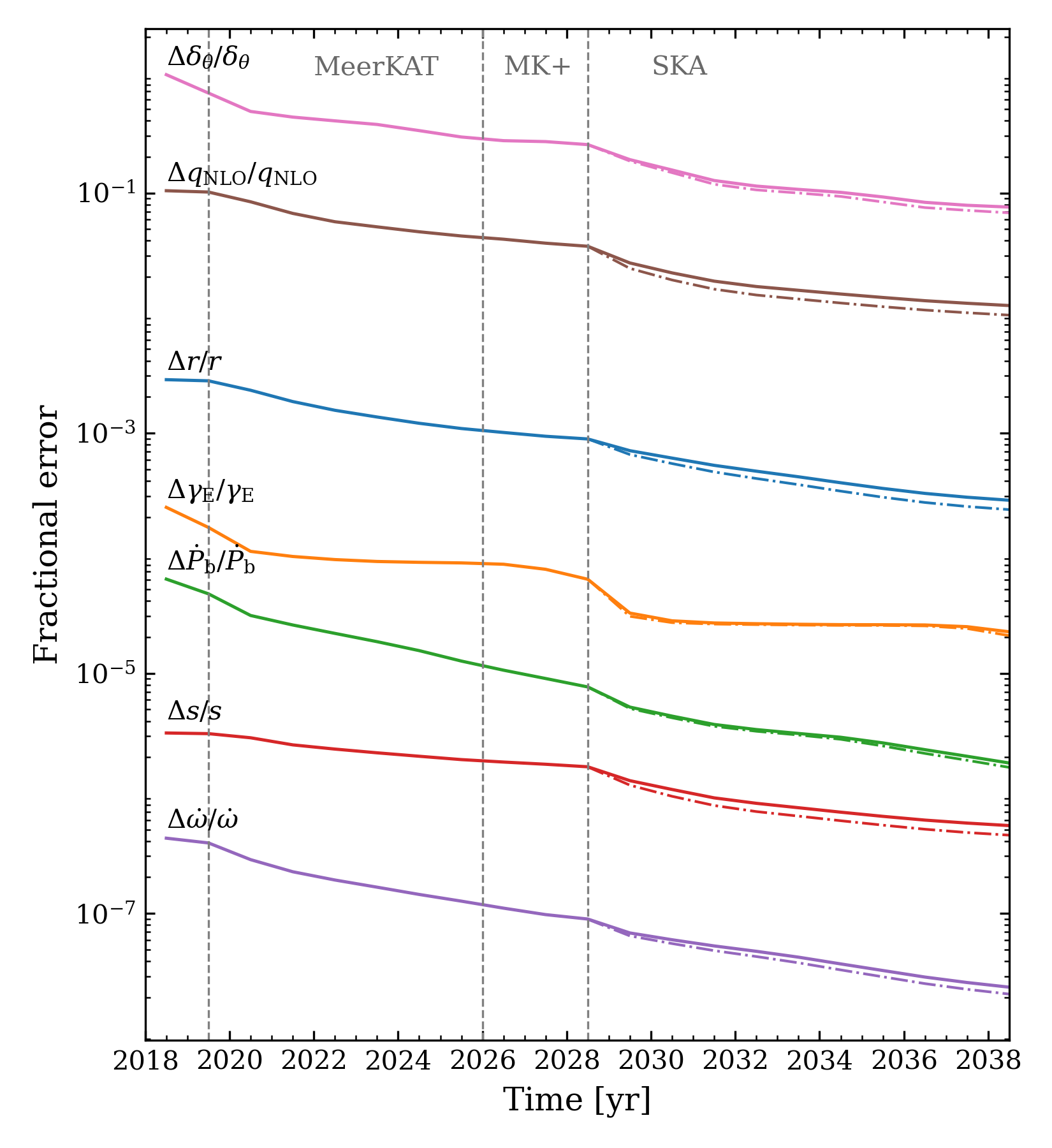}
    \caption{Predicted fractional errors of seven PK parameters (in log scale) with time. The vertical dashed lines mark the observing phase of MeerKAT, MeerKAT+(MK+), and SKA-Mid. From top to bottom are: orbital deformation parameter $\delta_\theta$ (pink), NLO factor for signal propagation $q_\mathrm{NLO}$ (brown), Shapiro delay range parameter $r$ (blue), Einstein delay amplitude $\gamma_\mathrm{E}$ (orange), orbital period derivative $\dot{P}_\mathrm{b}$ (green), Shapiro delay shape parameter $s$ (red), and relativistic advance of periastron $\dot{\omega}$ (purple). The solid lines represent simulations with MeerKAT, MeerKAT+(MK+) and SKA AA*, and the dash-dotted lines are those with SKA AA4 instead of AA*.}
    \label{fig:frac_error}
\end{figure}

Figure~\ref{fig:frac_error} shows the prediction of fractional errors of seven PK parameters with time, which are expected to be improved by more than an order of magnitude with continued observations using MeerKAT and the SKA. 
These will further improve the precision of gravity tests.

An interesting outcome in addition to this is the determination of neutron star moment of inertia (MOI) via relativistic spin-orbit coupling. 
The coupling of the spin of a rotating body and the orbital motion, which is also known as the Lense--Thirring (LT) effect \citep{LT1918}, leads to a precession of the orbit as a contribution to the observed rate of advanced periastron $\dot{\omega}$.
Therefore, the intrinsic contribution to $\dot{\omega}$ in the Double Pulsar system includes the 1PN and higher-order corrections due to 2PN effect and LT contribution of pulsar A \citep{Hu+2020,Kramer+2021PRX}:
\begin{align}
&\dot{\omega}^\mathrm{intr} = \dot{\omega}^\text{1PN}(m_\mathrm{A}, m_\mathrm{B}) +\dot{\omega}^\text{2PN} (m_\mathrm{A}, m_\mathrm{B}) +\dot{\omega}^\text{LT}_\mathrm{A}(m_\mathrm{A}, m_\mathrm{B}, I_\mathrm{A})  \,.\label{eq:omdot}
\end{align}
A simplified expression for each term is given in \cite{Hu2024Univ}. 
By 2038, the precision of the observed $\dot{\omega}$ is expected to reach $10^{-7}$ level (see Fig.~\ref{fig:frac_error}).
Other higher-order contributions such as the rotationally induced mass quadrupole moments of pulsars A and B to the orbital dynamics, LT contribution from pulsar B, and spin-spin coupling \citep{Barker1975} are still one or more orders of magnitudes below the measurement precision \citep{Hu+2020}. 
Apart from these intrinsic contributions, the observed value of periastron advance is also affected by the proper motion of the pulsar system, known as the ``Kopeikin term'' \citep{Kopeikin1996}: $\dot{\omega}^\mathrm{obs} = \dot{\omega}^\mathrm{intr} + \dot{\omega}^\mathrm{K}$. 
The magnitude of this contribution is also at the $10^{-7}$ level \citep{Hu+2020} and needs to be corrected.

If we know the masses of two pulsars $m_\mathrm{A}, m_\mathrm{B}$, the 1PN and 2PN can be calculated and deducted from the observed $\dot{\omega}^\mathrm{obs}$, and therefore one can measure the MOI of pulsar A, $I_\mathrm{A}$, from the LT contribution.
From Fig.~\ref{fig:frac_error} one can see that the two most precise PK parameters (after $\dot{\omega}$) for mass measurements are $s$ and $\dot{P}_\mathrm{b}$. 

In terms of $\dot{P}_\mathrm{b}$, the rate of change of orbital period, the situation is more complicated as the observed value not only contains intrinsic contributions from the pulsar system but is also affected by external effects from the Galaxy. 
For this particular system, the following contributions need to be consider:
\begin{align}
    \dot{P}_\mathrm{b}^\text{ obs}
    & = \dot{P}_\mathrm{b}^\text{ GW} (m_\mathrm{A}, m_\mathrm{B})+ \dot{P}_\mathrm{b}^{\,\dot{m}_\mathrm{A}} (m_\mathrm{A}, m_\mathrm{B}, I_\mathrm{A}) + \dot{P}_\mathrm{b}^{\text{ Gal}} + \dot{P}_\mathrm{b}^{\text{ Shk}} \,.
    \label{eq:pbdot}
\end{align}
The first two terms are intrinsic contributions from GW emission and spin down mass loss of pulsar A ($\dot{m}_\mathrm{A}$), respectively. 
The leading-order contribution from GW emission enters the equations of motion at 2.5 PN level (order $v^5/c^5$, quadrupolar GWs) \citep{Peters1963, Esposito1975, Wagoner1975}, whereas the NLO term occurs at 3.5 PN level (order $v^7/c^7$, octupolar GWs) \citep{BS89}, both are functions of two masses. 
According to the simulation, the precision of $\dot{P}_\mathrm{b}^\text{ obs}$ is expected to reach $10^{-3}\, \mathrm{fs/s}$ level by 2038, which is one order of magnitude smaller than the 3.5 PN GW contribution and the mass loss contribution of pulsar A, which  also depends on $I_\mathrm{A}$. 
Since the mass loss contribution scales inversely with the cube of the pulsar's spin period, and pulsar B rotates approximately 122 times more slowly than pulsar A, the mass loss contribution from B is negligible. 
The last two terms in Eq.~\eqref{eq:pbdot} are kinematic contributions due to the difference in Galactic acceleration (Gal) of the pulsar system and the Solar System \citep{DT1991ApJ},
\begin{equation}
    \left(\frac{\dot{P}_\mathrm{b}}{P_\mathrm{b}}\right)^{\text{Gal}} = \frac{1}{c} \vec{n}_\mathrm{10} \cdot (\vec{a}_\mathrm{1} - \vec{a}_\mathrm{0}) \,,
    \label{eq:Gal}
\end{equation}
and the proper motion of the pulsar, known as Shklovskii effect \citep[Shk,][]{Shk70}:
\begin{equation}
    \left(\frac{\dot{P}_\mathrm{b}}{P_\mathrm{b}}\right)^{\text{Shk}} = \frac{\mu^2 \, d}{c} \,.
\end{equation}
The quantity $c$ is the speed of light, and $\vec{n}_\mathrm{10}$ denotes the unit vector directed from the Solar System, index 0, toward the pulsar, index 1. 
Vectors $\vec{a}_\mathrm{0}$ and $\vec{a}_\mathrm{1}$ are the acceleration of the Solar System and the centre of mass of the binary system in the Galaxy, respectively.
$\mu$ denotes the proper motion of the pulsar as seen from the Solar System, and $d$ is the distance between the pulsar and the Sun.

Thus, the intrinsic orbital period decay can be extracted from the observed value as
\begin{align}
    \dot{P}_\mathrm{b}^\text{ intr}
    & = \dot{P}_\mathrm{b}^\text{ obs} -  \dot{P}_\mathrm{b}^{\text{ Gal}} - \dot{P}_\mathrm{b}^{\text{ Shk}}.
    \label{eq:intr}
\end{align}
A self-consistent approach to determining the MOI as developed in \cite{Hu+2020} is to account for these extrinsic contributions and mass loss of A in $\dot{P}_\mathrm{b}$ and solve for the three unknowns ($m_\mathrm{A}, m_\mathrm{B}$, $I_\mathrm{A}$) using $\dot{P}_\mathrm{b}^\text{ intr}(m_\mathrm{A}, m_\mathrm{B}, I_\mathrm{A})$, $\dot{\omega}^\text{ intr}(m_\mathrm{A}, m_\mathrm{B}, I_\mathrm{A})$, and $s\,(m_\mathrm{A}, m_\mathrm{B})$. 
However, to account for the total kinematic contribution $\dot{P}_\mathrm{b}^\text{D} = \dot{P}_\mathrm{b}^{\text{ Gal}} + \dot{P}_\mathrm{b}^{\text{ Shk}}$ require precise knowledge of pulsar distance and proper motion, as well as distance from the Sun to the Galactic centre $R_0$ and Galactic circular velocity at the location of the Sun $\Theta_0$.

\begin{figure}[ht]
    \vspace{-15pt}
    \centering
    \begin{subfigure}{0.67\textwidth}
    \includegraphics[width=\textwidth]{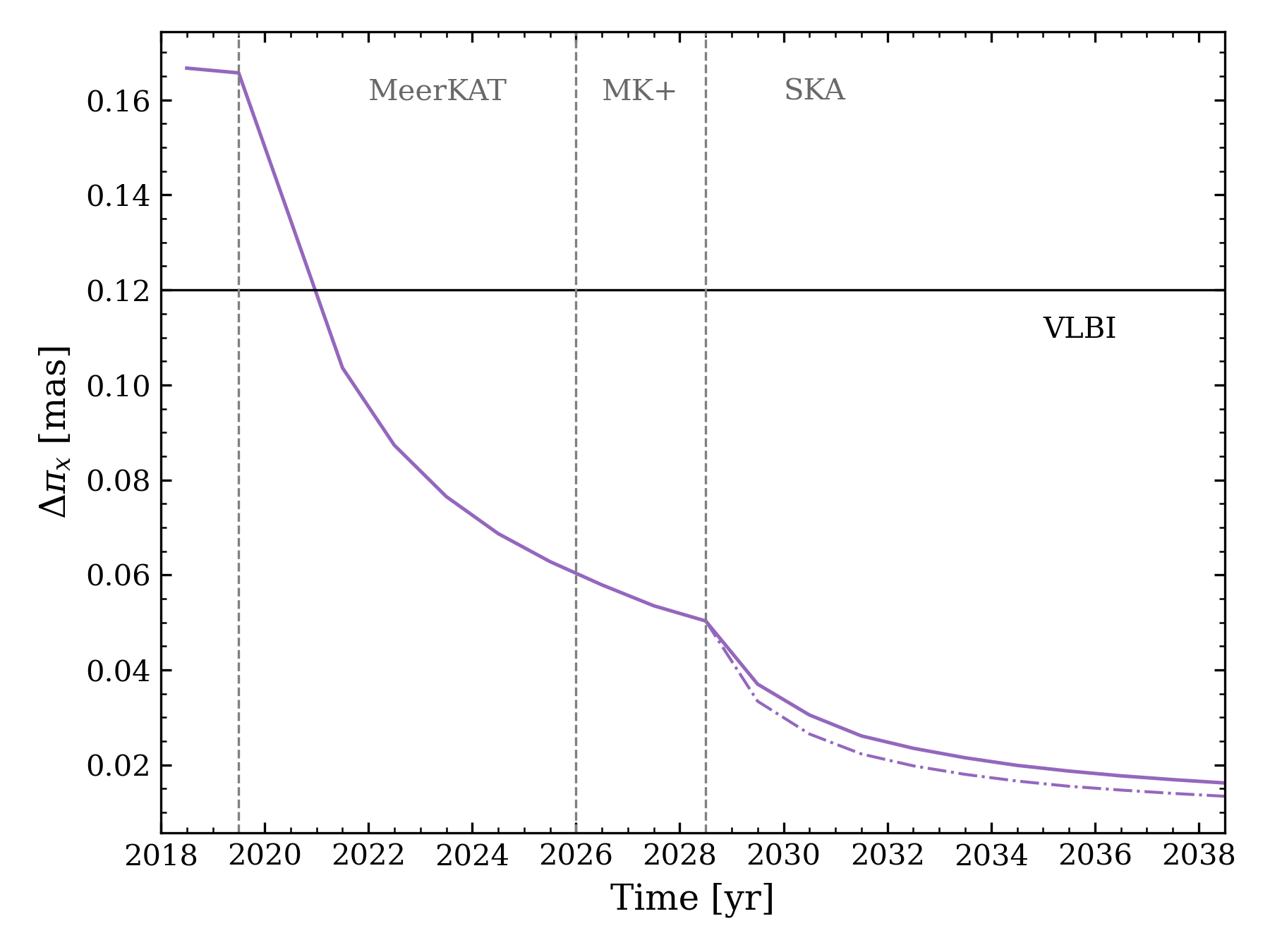}
    \caption{Timing parallax.}
    \label{fig:px}
    \end{subfigure}
    \begin{subfigure}{0.67\textwidth}
    \includegraphics[width=\textwidth]{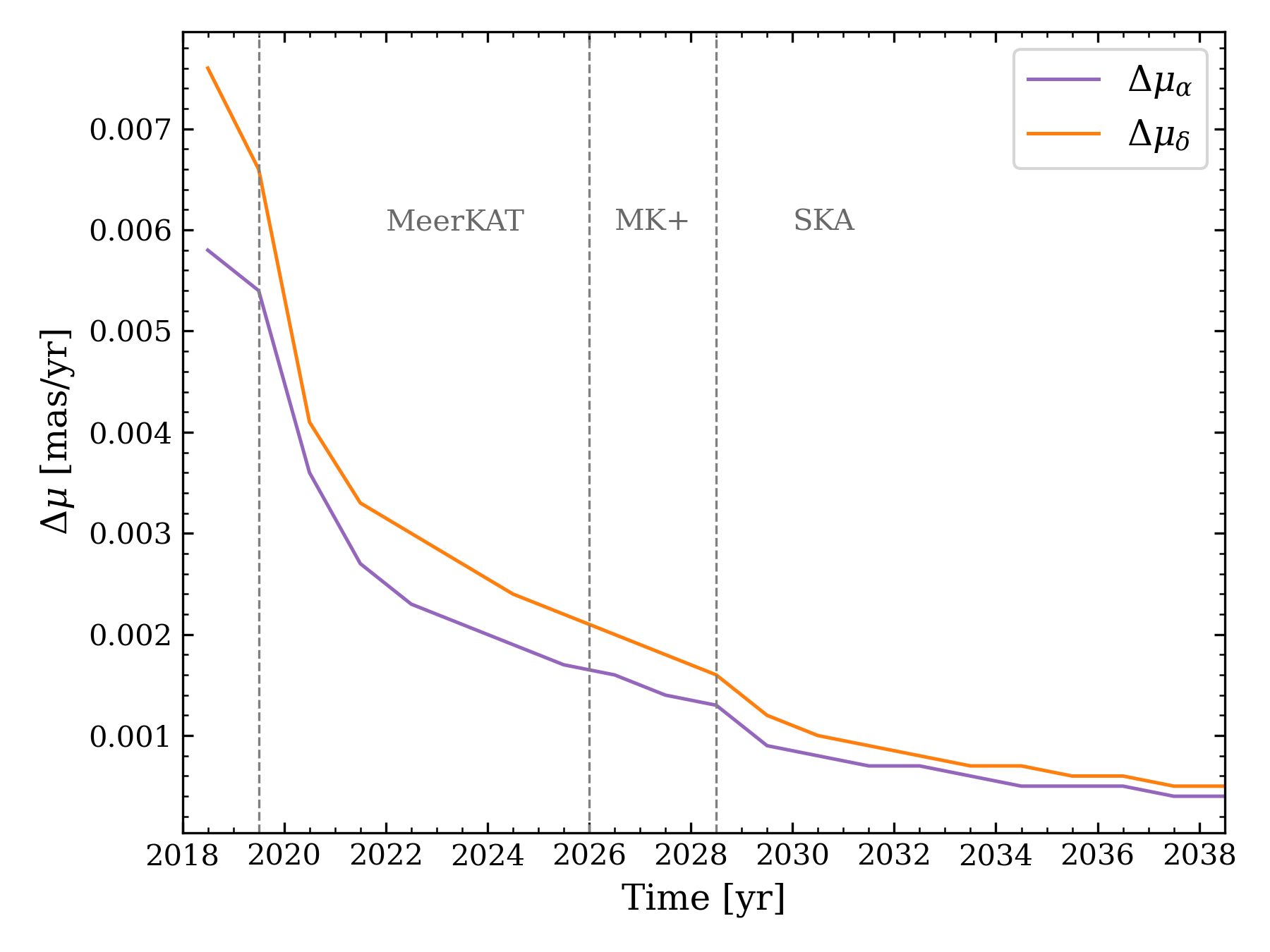}
    \caption{Proper motion.}
    \label{fig:pm}
    \end{subfigure}
    \caption{Predicted uncertainty of timing parallax $\pi_x$ and proper motion in right ascension $\mu_\alpha$ and declination $\mu_\delta$ using simulation data. The solid lines represent simulations with AA*, and the dash-dotted line represents simulations with AA4. For proper motion, there is not much difference between the two. The horizontal line in (a) indicates the VLBI parallax measured in \cite{Kramer+2021PRX}.}
    \label{fig:am}
    \vspace{-10pt}
\end{figure}

\begin{figure}[t]
    \vspace{-15pt}
    \centering
    \includegraphics[width=0.75\textwidth]{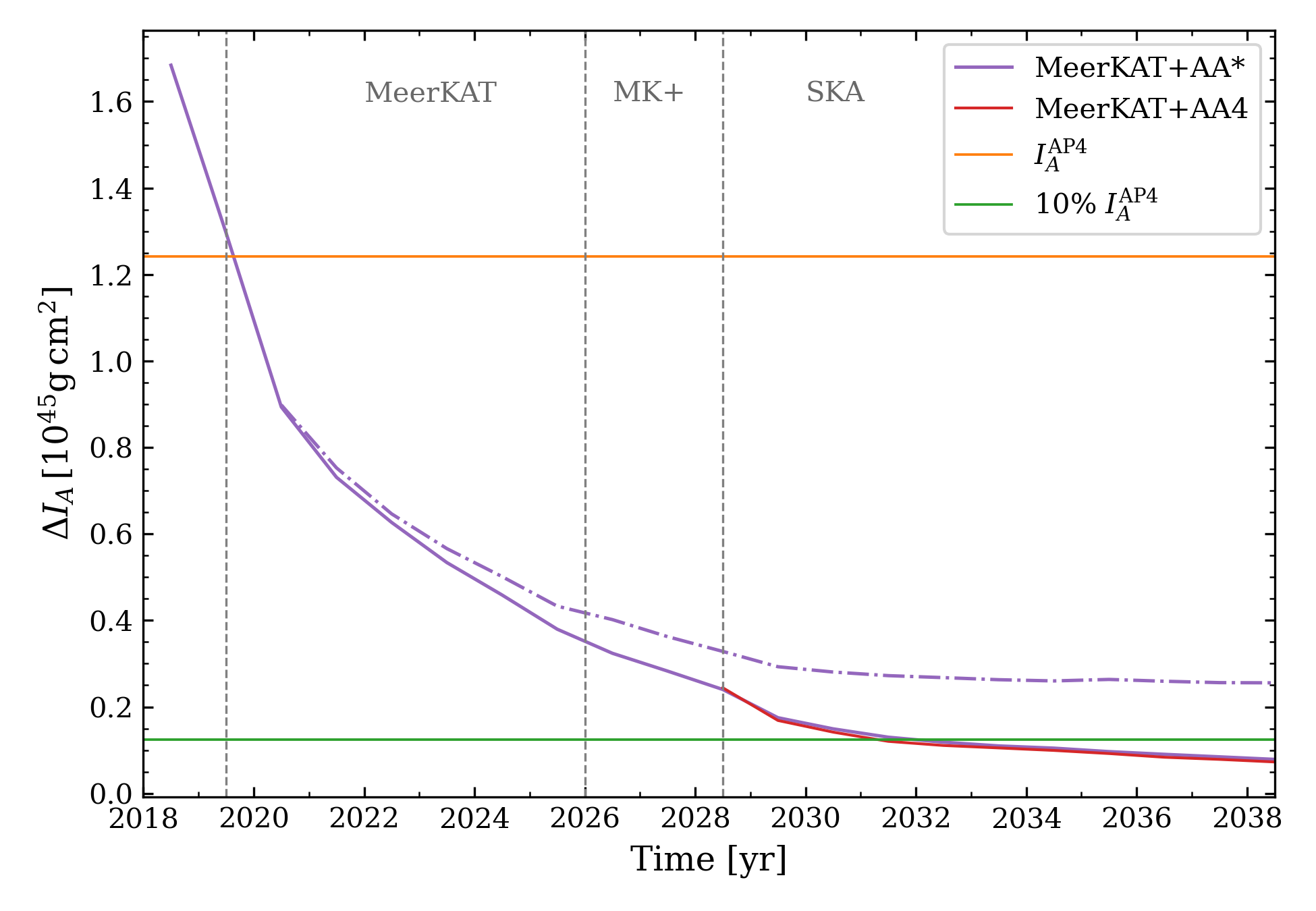}
    \caption{Predicted uncertainty of $I_\mathrm{A}$ as a function of time with simulated MeerKAT and SKA data. 
    The purple lines represent simulations with SKA AA*, and the red line indicate simulations with SKA AA4, which is marginally better than SKA AA*.
    The dash-dotted line adopts Galactic measurements in \protect\citet{Gravity2021, Guo+2021}, whereas the solid lines assumes no errors in the Galactic model. The orange horizontal line indicates the theoretical value of the MOI of the assumed EOS AP4, and the green line represents 10\% of the theoretical value.}
    \label{fig:IA}
\end{figure}

For relatively nearby pulsars, astrometric parameters can be measured using either pulsar timing or Very Long Baseline Interferometry (VLBI). 
Details of these two methods are discussed later in Section~\ref{sec:astrometry}. 
The recent VLBI parallax measurement is $1.30^{+0.13}_{-0.11}$~mas, which implies a pulsar distance of $770\pm 70$~pc \citep{Kramer+2021PRX}. 
According to the simulation, the precision of timing parallax $\pi_x$ is expected to improve with continuous observations with MeerKAT and the SKA, which is shown in Fig.~\ref{fig:px}.
In this simulation, red noise and variations in dispersion measure (DM) are not modelled. 
Under this circumstance, the accuracy of the timing parallax is expected to exceed the accuracy of previous VLBI measurements from 2021.\footnote{In practice, the presence of red and DM noise can enlarge the uncertainty of timing parallax (and proper motion) by about a factor of three (Hu et al. in prep.), delaying this outperformance to 2030. However, a better VLBI measurement is expected in the near future anyway, therefore these uncertainties are adopted for the MOI test.} 
For the distance used in the calculation of kinematic contributions, the uncertainty of the VLBI parallax is used until the uncertainty of the timing parallax is better. 
The simulation also predicts the uncertainty of proper motion in right ascension ($\mu_\alpha$) and declination ($\mu_\delta$) with time, as is shown in Fig.~\ref{fig:pm}. 
These values are much more precise than from VLBI measurements and are adopted in the calculation of kinematic contributions in order to deliver a more realistic estimation of the precision of $\dot{P}_\mathrm{b}^\text{D}$.
The Galactic measurements used in the calculation are: $R_0 = 8.275(34)$~kpc \citep{Gravity2021} and $\Theta_0 = 240.5(41)\, \mathrm{km\, s^{-1}}$ \citep{Guo+2021}. 

After removing the extrinsic contributions and performing the $\dot{P}_\mathrm{b}^\text{ intr}$--$\dot{\omega}^\text{ intr}$--$s$ test, $I_\mathrm{A}$ is measured and its uncertainty is shown in Fig.~\ref{fig:IA}. 
The dash-dotted line shows the uncertainty of $I_\mathrm{A}$ based on the current Galactic measurements, which decreases with time and levels off from 2030 onwards ($\sim23\%$ by 2030 and $\sim20\%$ by 2038, both with 68\% confidence).
This indicates that the measurements of $R_0$ and $\Theta_0$ will limit the accuracy of the $I_\mathrm{A}$ measurement, especially $\Theta_0$ which has a larger uncertainty. 
However, these measurements are expected to be much improved in the near future with, for example, \emph{Gaia} data \citep{Gaia2021Klioner}.

Assuming $R_0$ and $\Theta_0$ will be much improved in the future and not limit the uncertainty of $I_\mathrm{A}$, the solid lines (purple: AA*, red: AA4) indicate the result under the assumption that there is no error in these Galactic parameters.\footnote{A similar result is obtained if assuming $\Theta_0$ is significantly improved, e.g. by a factor of 100.}
The accuracy of $I_\mathrm{A}$ is expected to reach $\sim 10\%$ by 2030 and $\sim 5\%$ by 2038 (68\% confidence). 
The former is consistent with the previous simulation \citep{Hu+2020} despite the delay and reduced capability of MeerKAT+ and the SKA. 
This is because the timing accuracy of the UHF band data (not included in previous simulations) is 1.75 times that of the L band data, helping to compensate for this loss. 

This long-awaited MOI measurement will provide important additional constraints on the EOS of matter at supranuclear densities. 
Combining this with future neutron star mergers observed by GW detectors, X-ray observations, and advances in nuclear theory, the EOS could even be identified. 
This will advance our understanding of the properties of supranuclear matter, the composition of neutron star cores, and the mass boundary between neutron stars and black holes.

Furthermore, it has been shown that if the neutron star EOS can be well determined in the near future, it could in turn be used to test the Lense--Thirring effect ($\sim7\%$ by 2030) and provide valuable insights for testing alternative theories, or allow for a test of the 3.5 PN GW contribution \citep{Hu+2020}.

\section{Testing dipolar GWs from scalar-tensor theories using pulsar--white dwarf system PSR~J2222$-$0137}
\label{sec4}

The lowest multipole for the generation of GWs predicted in GR is the mass quadrupole. 
The lack of any time-dependent mass monopole or mass dipole is strongly linked to the validity of the Strong Equivalence Principle (SEP) and the resulting effacement of a body's internal structure \citep{Damour1987,Will2018book}.
However, in alternative gravity theories that violate the SEP, such as scalar-tensor gravity, dipolar gravitational radiation is generally predicted in binary systems \citep{Eardley1975, Will1977} due to the presence of additional gravitational fields (e.g. a scalar field). 
This leads to an additional dipolar GW contribution to the rate of change of orbital period: $\dot{P}_\mathrm{b}^\mathrm{Dipole} \propto k_\mathrm{D} \mathcal{S}^2\, c^{-3}$, which already enters the equations of motion at the 1.5 PN order. 
The quantity $k_\mathrm{D}$ is a theory-dependent constant that quantifies the dipolar self-gravity contribution, and $\mathcal{S} = s_\mathrm{p} - s_\mathrm{c}$ is the difference in the ``sensitivities'' of the pulsar, $s_\mathrm{p}$, and the companion, $s_\mathrm{c}$ \citep{Will1993}. 
The value of $s_\mathrm{p,c}$ depends on the theory of gravity, the exact form of the EOS, and the mass of the body. 
The sensitivity of neutron stars is generally of the order of 0.15, whereas the sensitivity of white dwarfs (WDs) is negligible: $s_\mathrm{WD} \lesssim 10^{-3}$.
Therefore, binary pulsar experiments are generally sensitive to the dipolar radiation, especially those with significant asymmetry in the sensitivities, such as pulsar--WD systems. 
For instance, the radiative tests with the pulsar--WD system PSR~J1738+0333 \citep{Freire+2012MNRAS} has placed stringent limits on dipolar GW emission. 

\begin{figure}
    \vspace{-10pt}
    \centering
    \includegraphics[width=\linewidth]{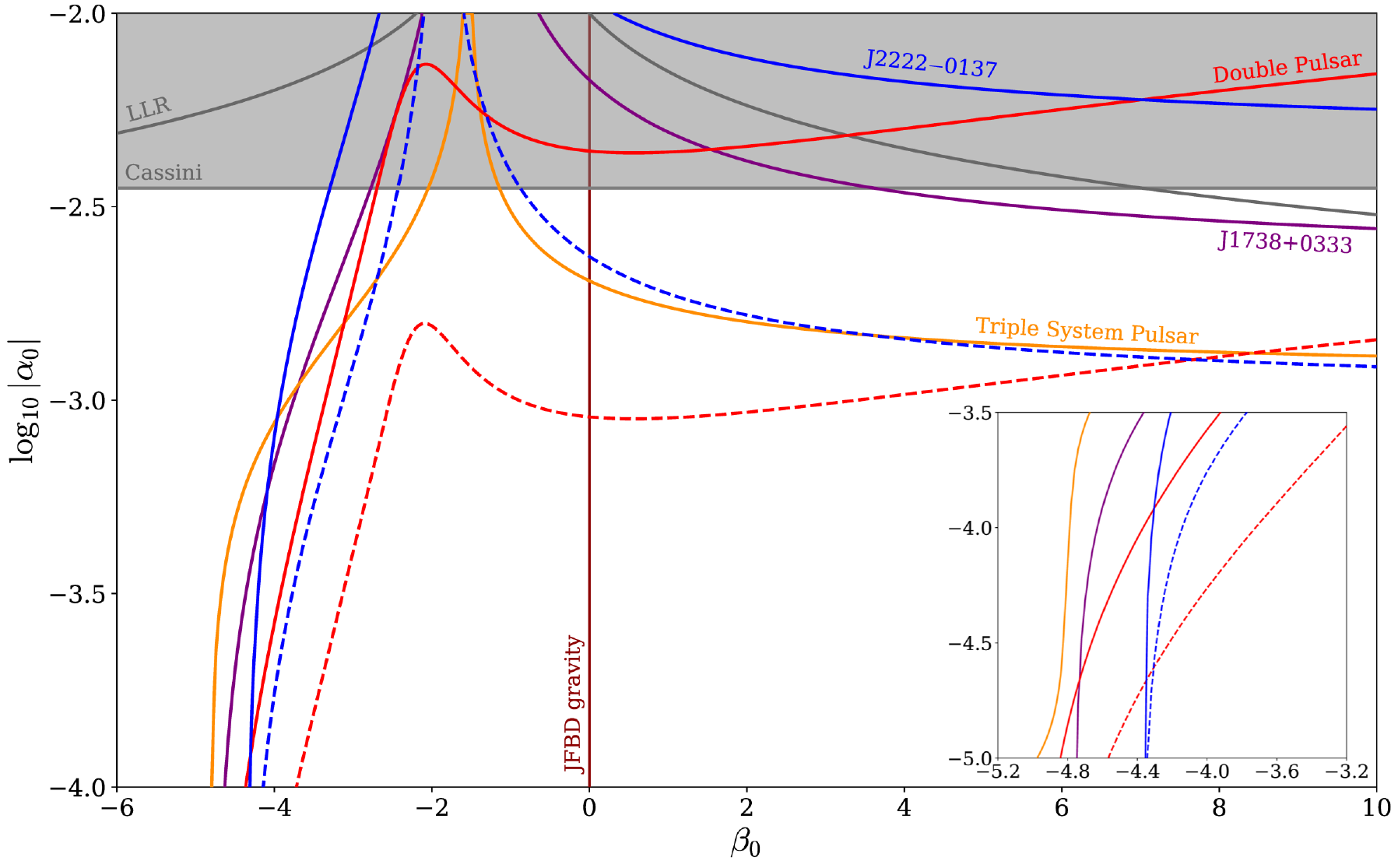}
    \caption{Constraints on the DEF gravity parameter space from different experiments to date (95\% confidence, solid lines): Shapiro delay with the Cassini spacecraft \citep{Bertotti2003_Cassini}, dipolar radiation with J1738+0333 and J2222$-$0137, Nordtvedt effect from Lunar Laser Ranging \citep[LLR,][]{Biskupek2021Univ} and the triple system pulsar \citep{Voisin2020A&A},
    as well as the Double Pulsar \citep{Kramer+2021PRX}. Areas above a curve are excluded by the corresponding experiment. Pulsar curves are computed with a relatively stiff EOS MPA1 \citep{Lattimer2001ApJ}. Predicted constraints based on simulated data of the Double Pulsar and J2222$-$0137 are shown as dashed lines. The insert shows the constraints in the area where $\alpha_0 <10^{-3.5}$. Courtesy of Norbert Wex.}
    \label{fig:DEF}
\end{figure}

Another binary system that is suitable for constraining alternative theories is  PSR~J2222$-$0137, a mildly recycled pulsar with a spin period of 32.8~ms and an orbital period of 2.45~days \citep{Boyles+2013ApJ}. 
The precise measurement of $\dot{P}_\mathrm{b}$ and the most precise pulsar distance measurement from VLBI observations allowed the extrinsic kinematic contributions to be corrected and, leading to a $2\sigma$ measurement of the GW damping at 2.5 PN order \citep{Guo+2021}. 
The relatively large mass of the pulsar, $1.831\pm 0.010\,\mathrm{M}_\odot$, makes it a particularly interesting system for certain non-linear deviations from GR in the strong fields and closing the mass gap of spontaneous scalarisation \citep{ZhaoJuejie2022CQG} in Damour--Esposito-Far\`{e}se (DEF) gravity. 

The DEF gravity is a mono-scalar–tensor theory that contains a massless scalar field in addition to the spacetime metric \citep{DEF1992, DEF1993, DEF1996}. 
The two parameters $\alpha_0$ and $\beta_0$ define the two-dimensional parameter space of this class of theories, with $\alpha_0=\beta_0=0$ for GR and $\beta_0=0$ and $\alpha_0^2 = (2\omega_\mathrm{BD}+3)^{-1}$ for the Jordan-Fierz-Brans-Dicke (JFBD) gravity \citep{Jordan1955,Fierz1956,Jordan1959,BD1961}, where $\omega_\mathrm{BD}$ is the Brans--Dicke parameter. 
Figure~\ref{fig:DEF} illustrates constraints in the DEF gravity parameter space ($\alpha_0, \beta_0$) obtained from various Solar System and pulsar experiments. 
The solid lines show 95\% constraints from current experiments, where area above a curve is excluded by the given experiment. 
For instance, Shapiro delay with Cassini gives the best limit for $-2.0 \lesssim \beta_0 \lesssim -1.1$.
In particular, the triple star system PSR~J0337+1715 \citep{Archibald2018Natur,Voisin2020A&A}, which contains two WDs, contributes currently the best limits in the region $\beta_0 \gtrsim -1.1$ and $-3.1 \lesssim \beta_0 \lesssim -2.0$. 
Despite the less asymmetry in the Double Pulsar, it contributes the most in the region $-4.3\lesssim \beta_0 \lesssim -3.1$. 
And PSR~J2222$-$0137 places the most stringent constraint in the highly non-linear region $\beta_0 \lesssim -4.3$. 
Based on the simulated SKA data of the Double Pulsar in Section~\ref{sec32}, it is expected to give much tighter constraint as indicated by the red dashed line.
Due to the complexity of the orbital motion of the triple system, predicting its future capacity through simulations is not feasible, whereas PSR~J2222$-$0137 is still a very promising system given its precisely known distance. 

In my previous simulations, it has been shown that continuous observations of PSR~J2222$-$0137 using FAST and MeerKAT will greatly improve the limits on DEF gravity parameters, especially in the strongly non-linear part of the parameter space
\citep{Batrakov+2023}.
Here as an update, I perform similar simulations for MeerKAT (starting from 2021), MeerKAT+, and the SKA AA4, as well as for FAST (covering 2021--2038). 
In addition, to improve the angular resolution of FAST, a Core Array extension has been proposed and started construction last year \citep{FASTcore2024}. 
The FAST Core Array integrates 24 secondary 40-m antennas implanted within 5~km of the FAST site and is estimated to be completed and put into operation in 2027.\footnote{See \url{https://english.www.gov.cn/news/202409/25/content_WS66f3a4c1c6d0868f4e8eb3fb.html}} 
The effective diameter of this fully steerable 40-m array alone is about 196~m, comparable to the SKA AA4. By integrating FAST into this synthesis array, the total effective diameter amounts to 358~m, exceeding the capabilities of other next-generation arrays such as the SKA and the next-generation Very Large Array. 
The project is expected to eventually expand to integrate a total of 64 secondary 40-m antennas within 30~km radio quiet zone, with stand-alone and total effective diameters of 320~m and 439~m, respectively. 

\begin{table}[t]
{\renewcommand{\arraystretch}{1.5}
\centering
\caption{\small Summary of simulation setup for PSR~J2222$-$0137 based on simulations in \cite{Batrakov+2023}. 
The TOA uncertainty are estimated from L-band observations and scaled to a sub-integration of 15~min over the full bandwidth. 
A full orbit (2.45~days) is covered every year and the observation is divided into monthly cadence. The effective diameter (without and with FAST) and TOA uncertainty for FAST Core Arrays are listed as a reference, but are not included in the simulations.}
\label{tab:sim2}
\begin{tabular}{@{}lcccccc@{}}
\toprule
Telescope & MeerKAT & MeerKAT+ & 
SKA AA4 & FAST & Core Array I & Core Array II\\ 
\midrule
Effective diameter [m] & 108  & 111 & 204 & 300 & 196$|$358 & 320$|$439\\ 
Observing period [yr] & \makecell{2021/Q1-\\2025/Q4} & \makecell{2026/Q1-\\2028/Q2} & \makecell{2028/Q3-\\2038/Q2} & \makecell{2021/Q1-\\2038/Q3} & - & - \\ 
TOA uncertainty [$\mu\rm{s}$] & 0.84 &  0.79 & 
0.24 & 0.12 & 0.10 & 0.08\\ 
\botrule
\end{tabular}}
\end{table}

\begin{table}[t]
\vspace{-15pt}
\centering
{\renewcommand{\arraystretch}{1.3}
\caption{\small Expected improvements in the precision of timing parameters of PSR~J2222$-$0137 in 2038 as compared to astrometric parameters measured in \cite{Ding+2024} and PK parameters measured in \cite{Guo+2021}. Case 1: MeerKAT+SKA(AA4). Case 2: MeerKAT+SKA(AA4)+FAST. }
\vspace{5pt}
\begin{tabular}{cccc}
    \toprule
       & MK+SKA & MK+SKA+FAST   &  Fractional error of Case 2 \\ \midrule
        $\mu_\alpha$ & 2.8 & 8.3 & $1.3\times10^{-5}$ \\ 
        $\mu_\delta$ & 2.8 & 8.6 & $2.5\times10^{-4}$ \\ 
        $\pi_c$ & 1.2 & 3.1 & $1.1\times10^{-3}$\\ 
        $\dot{\omega}$ & 11.5 & 36.1  & $1.4\times10^{-4}$ \\ 
        $\dot{P}_\mathrm{b}$ & 29.3 & 56.0 & $9.2\times10^{-3}$ * \\ 
        $s$  & 5.3 & 14.4 & $8.2\times10^{-6}$ \\ 
        $m_\mathrm{c}$ &  4.9 & 13.3  & $5.2\times10^{-4}$ \\ 
        $\dot{x}$ & 10.5 & 31.8  & $1.9\times10^{-3}$ \\ 
        \botrule
\end{tabular}
\footnotetext{* This is the ratio of the observed error to the intrinsic value of $\dot{P}_\mathrm{b}$.}
\label{tab:J2222}
}
\end{table}

\begin{figure}[ht]
    \vspace{-10pt}
    \centering
    \includegraphics[width=0.7\linewidth]{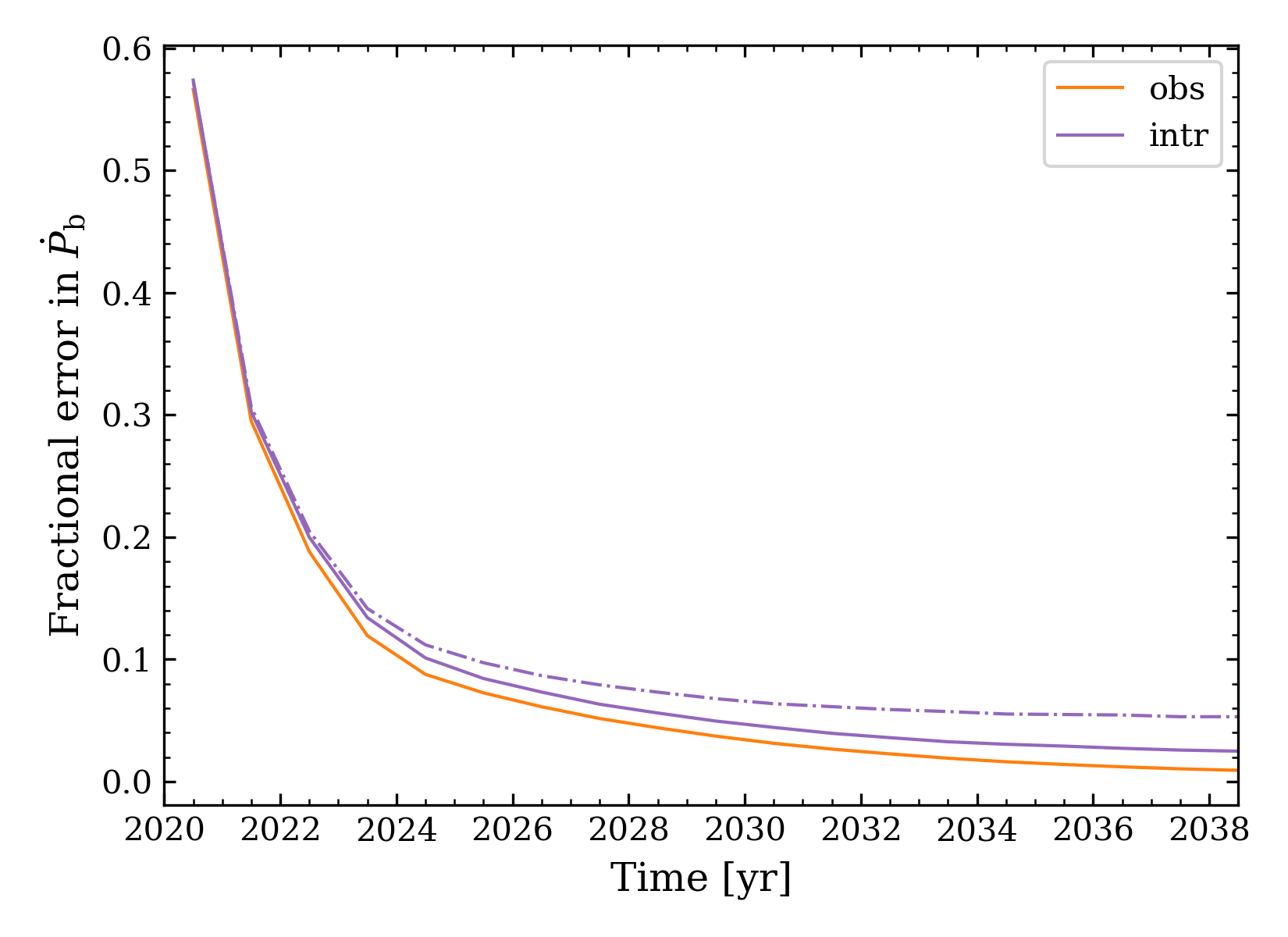}
    \caption{Predicted fractional error of $\dot{P}_\mathrm{b}$ based on simulations of PSR~J2222$-$0137 using  MeerKAT+SKA(AA4)+FAST. The orange line represents the ratio of the observed error to $\dot{P}_\mathrm{b}^\mathrm{\,intr}$, whereas the purple lines represent fractional error of $\dot{P}_\mathrm{b}^\mathrm{\,intr}$ after accounting for external uncertainties and an improving timing parallax. The dash-dotted line assumes the same Galactic model as in Fig.~\ref{fig:IA}, while the solid purple line assumes a 100-fold improvement in the precision of $\Theta_0$.}
    \label{fig:dPBdot}
    \vspace{-10pt}
\end{figure}

The simulation specification for PSR~J2222$-$0137 is summarised in Table~\ref{tab:sim2}, and the improvements in timing parameters by 2038 compared to the recent results are listed in Table~\ref{tab:J2222}. 
One can see that the addition of FAST improves the precision of timing parameters by approximately three times compared to MeerKAT and the SKA, especially for $\dot{P}_\mathrm{b}$, proper motion, and parallax, which are required for testing dipolar GWs. 
Compared to \cite{Guo+2021}, the observed error of $\dot{P}_\mathrm{b}$ will be improved by more than 50 times. 
Figure~\ref{fig:dPBdot} shows the improvement in the fractional error of $\dot{P}_\mathrm{b}^\mathrm{\,intr}$ from 2021 to 2038, with the observed error in orange and the intrinsic error in purple (after accounting for the uncertainties in distance, proper motion, and the Galactic model). 
Considering that future Galactic models will be much improved, the solid purple line suggests that a 40-$\sigma$ test of GW will be possible by 2038, a 20-fold improvement over \cite{Guo+2021}, thereby a much improved constraint on the DEF parameter space, as indicated by the blue dashed line in Fig.~\ref{fig:DEF}. 
It is expected to surpass the Double Pulsar (red dashed line) in the region where $\beta_0<-4.3$ and $\beta_0>7.7$. 
Other binary systems, such as the asymmetric double neutron star system PSR~J1913+1102 \citep{Ferdman2020Natur}, are potentially good laboratories for such tests.
Future observations using FAST and the FAST Core Array will play a key role in testing the theories of gravity. 

\section{Nanohertz GW breakthrough using PTAs}\label{PTA}

In June 2023, a major breakthrough occurred in the search for low-frequency GWs. 
Our work with the European Pulsar Timing Array (EPTA), alongside teams in China (CPTA), Australia (PPTA), and North America (NANOGrav), revealed compelling evidence for a stochastic GW background (GWB) in the nanohertz frequency range \citep{EPTA2023A&AIII,CPTA2023RAA,PPTA2023ApJ,NanoGrav2023ApJ}. 
This signal matches what we expect from pairs of supermassive black holes slowly spiralling toward collision throughout the cosmos. 
The distinctive fingerprint of these GWs, known as the Hellings-Downs (HD) correlation, was independently confirmed by the MPTA team using MeerKAT data \citep{MPTA2025}.

At the heart of this discovery are decades of patient data collection. 
The EPTA Data Release 2 (DR2) represents a significant milestone in this endeavour, comprising nearly 24.7 years of observations from 25 MSPs. 
These data were collected using five major European radio telescopes: the Effelsberg 100-m Radio Telescope in Germany, the 76-m Lovell Telescope in the United Kingdom, the Nançay Radio Telescope (NRT) in France (which has a collecting area equivalent to a 94-m diameter dish), the Westerbork Synthesis Radio Telescope (WSRT) in the Netherlands (an array of fourteen 25-m antennas equivalent to a 94-m single dish), and the 64-m Sardinia Radio Telescope (SRT) in Italy. 
These five telescopes also regularly operated as the Large European Array for Pulsars (LEAP), which performs simultaneous phased-array observations, achieving an effective diameter of up to 194~m \citep{Bassa2016LEAP}. 
This unique observing mode not only provides the highest-precision data in EPTA DR2, but also further improves pulsar timing accuracy by cross-referencing the delays between telescopes to obtain the correct clock files \citep{Hu2023PhDT}.

The earliest data in EPTA DR2 date back to observations made with the Effelsberg Berkeley Pulsar Processor (EBPP) in 1996, with most pulsars in the sample having over 15 years of timing data. 
The dataset is divided into two versions: ``DR2full'', which includes all available EPTA DR2 data, and ``DR2new'', which consists of a more recent 10.3-yr subset obtained using modern backends.

\begin{figure}[t]
    \vspace{-10pt}
    \centering
    \includegraphics[width=0.7\textwidth]{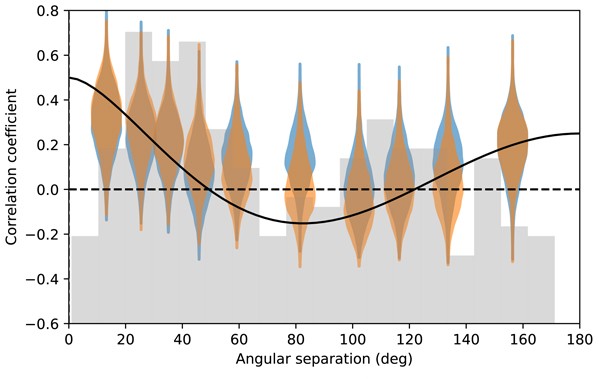}
    \caption{Constraints on the HD correlation (or overlap reduction function, ORF) from the Bayesian analysis, shown as violins of the posterior of the correlation coefficients averaged at ten bins of angular separation.
    The black line denotes the HD curve based on theoretical expectation of a GWB signal, and the grey histogram represents the distribution of pulsar pairs at different separation. Adapted from \cite{EPTA2023A&AIII} under CC BY 4.0.}
    \label{fig:epta2}
\end{figure}

\begin{figure}[t]
    \vspace{-10pt}
    \centering
    \includegraphics[width=\textwidth]{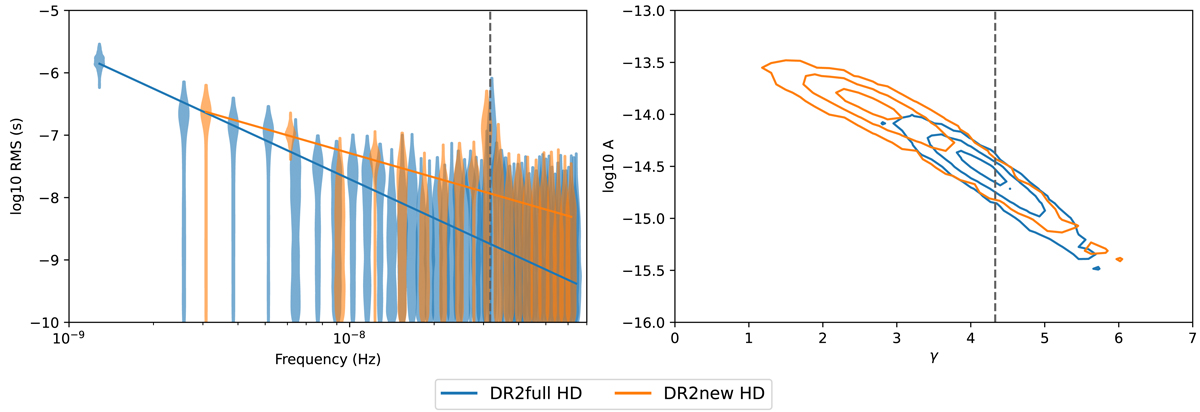}
    \caption{Spectral properties of a common red signal assuming HD correlation. 
    The left panel displays the free spectrum, representing independent measurements of common power across each frequency bin for DR2full (blue) and DR2new (orange) dataset. The solid lines indicate the best-fit power law models for the GWB, and the vertical dashed line indicates the position of $f=1 ~\mathrm{yr}^{-1}$.
    The right panel illustrates the 1/2/3$\sigma$ contour of the 2D posterior distribution for the amplitude and spectral index, obtained by modelling the spectrum with a power law. The vertical dashed line denotes the theoretical value $\gamma=13/3$. Reproduced from \cite{EPTA2023A&AIII} under CC BY 4.0.}
    \label{fig:epta1}
\end{figure}

The PTA analysis focuses on pulsar timing residuals, which track tiny deviations in pulse arrival times that could indicate the influence of GWs. 
A key part of the study involves identifying a common-spectrum process, meaning a low-frequency signal present in all pulsars that cannot be attributed to individual pulsar noise. 
If a stochastic GWB exists, it induces correlated delays in the arrival times of pulses from different pulsars, with the correlation strength depending on their angular separation. 
This correlation is known as the Hellings-Downs (HD) correlation \citep{HD1983}, where pulsars close together in the sky show positive timing correlations, while those at large separations ($\sim 90\si{\degree}$ apart) exhibit negative correlations. 
This expected correlation of a GWB signal is shown as the black curve in Fig.~\ref{fig:epta2}. 
The violins represent the posterior of the correlation coefficients from the Bayesian search for the HD correlation, with ten angularly spaced bins, each comprising 30 pulsar pairs. 
Comparing to DR2full (blue), the DR2new (orange) constraints are much more consistent with the expected HD correlation. 
With the 10.3-yr subset, the statistical analysis reveals strong evidence for a GWB, with a Bayes factor of 60, suggesting a very low probability (approximately 0.1\%) that the signal is due to random noise. 

To characterise this signal, the power spectrum was also analysed, employing both the free-spectrum method, which estimate power independently at each frequency, and power-law modelling, which assumes a spectral shape consistent with a GWB from SMBHBs. 
The spectrum is shown in the left panel of Fig.~\ref{fig:epta1}, while the amplitude $A$ and spectral index $\gamma$ is shown on the right. 
Results for DR2full are shown in blue and results for DR2new are shown in orange. 
The long data baseline was crucial for the DR2full to reach a spectral consistent with the expected value ($\gamma=13/3$) for a GWB from circular SMBHBs (dashed line on the right panel).

The EPTA results are highly consistent with those from other major PTAs, all of which report similar findings in their independent datasets. 
The next step in confirming the existence of the nanohertz GWB involves a global collaboration through the International Pulsar Timing Array \citep[IPTA,][]{IPTA2021}, which will combine data from all PTAs to improve sensitivity and refine the signal. 
If the GWB signal can be confirmed, it promises to improve our understanding of galaxy formation and evolution, and opens up the possibility of revolutionary advances in the study of the early Universe (inflation, phase transitions, cosmic string, etc.), dark matter, and modified gravity \citep{Burke-Spolaor2019}, and even identifying nearby individual SMBHBs \citep{Mingarelli2019}.

\section{Current limitations in pulsar experiments}\label{limit}

Despite the current advances in gravitational experiments with radio pulsars, there are challenges that limit our ability to go beyond the current precision of tests and understanding of gravity. 

\begin{figure}[t]
    \vspace{-15pt}
    \centering
    \includegraphics[width=0.78\textwidth]{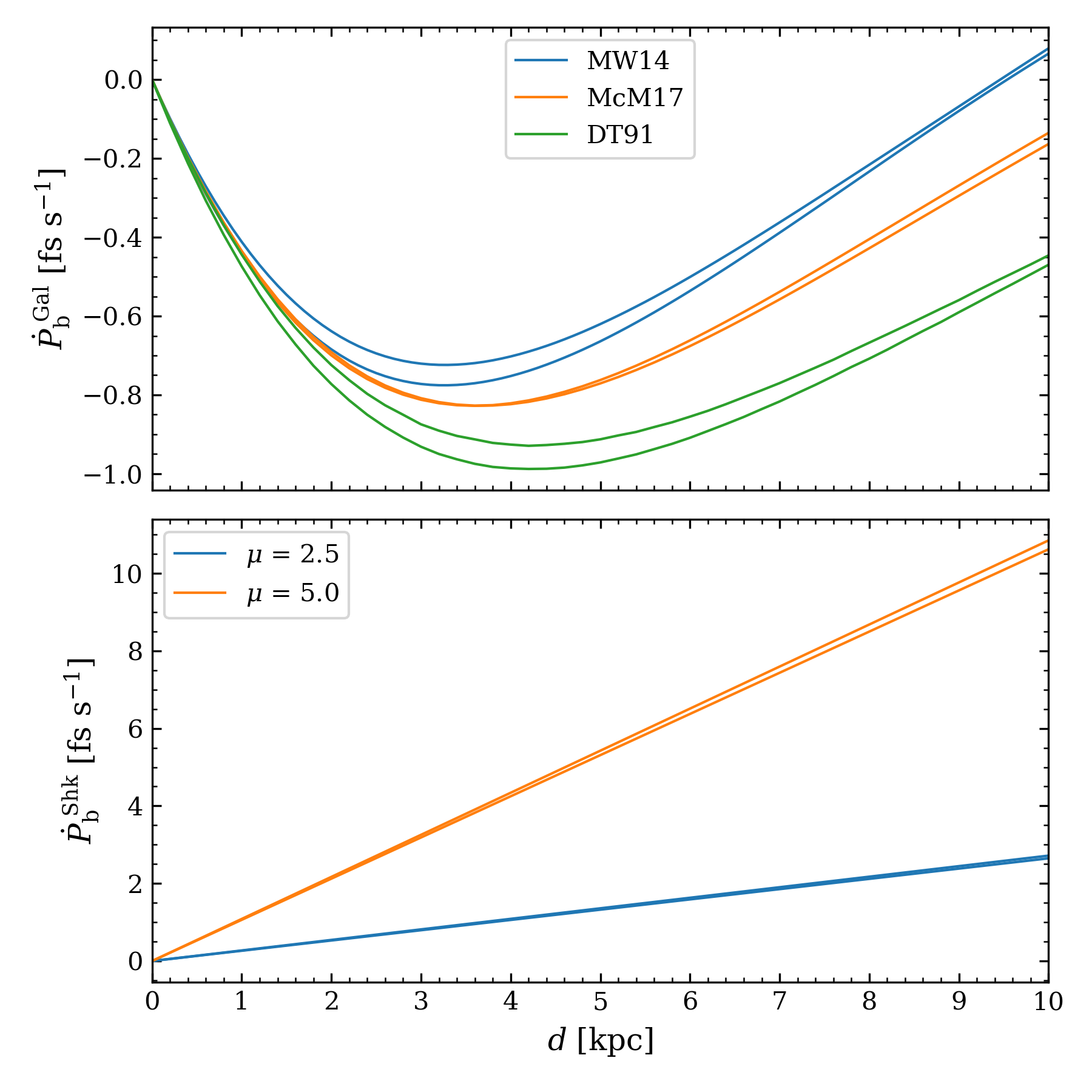}
    \caption{Galactic (top) and Shklovskii (bottom) contributions to the orbital period derivative $\dot{P}_\mathrm{b}$ as a function of distance using different Galactic models and proper motion $\mu$. Dual lines indicate $1\sigma$ error.}
    \label{fig:Gal}
    \vspace{-10pt}
\end{figure}

For example, a key parameter in the binary pulsar experiments is the intrinsic derivative of orbital period $\pbdot^\mathrm{intr}$. 
As shown in Eq.~\eqref{eq:pbdot}, the observed decay in the orbital period $\dot{P}_\mathrm{b}$ not only results from the energy loss due to GW emission, but also contains contributions from the relative acceleration between the pulsar system and the Solar System Barycentre (SSB), requiring precise knowledge on the pulsar distance, its transverse velocities (proper motion), and the gravitational potential of the Galaxy. 
Figure~\ref{fig:Gal} shows the Galactic and Shklovskii contributions at different distances. 
To demonstrate the impact of the usage of different Galactic potential models, the top panel displays a comparison of three widely used Galactic potential models (in the direction of the Double Pulsar) in binary pulsar experiments. Blue and orange represent the ``MWpotential2014'' (MW14) model and the ``McMillan17'' (McM17) model \citep{McMillan2017}, respectively, through the \textsc{galpy} package \citep{Bovy2015galpy}. 
While the green (DT91) denotes the analytical approximation of Eq.~\eqref{eq:Gal} with vertical acceleration \cite[see][]{DT1991ApJ,Lazaridis09, Hu+2020}. 
However, this approximation assumes a linear rotation curve and therefore only applies to pulsars with Galactic radii similar to the Sun. 
The difference between these three models is below $1\sigma$ for distance $d<1$~kpc, but becomes increasingly large for large distances. 
As many binary pulsars are relatively distant, such as the HT pulsar B1913+16 and the most compact double neutron star system PSR~J1946+2052 \citep{Stovall+2018}, a precise knowledge of the Galactic potential model, as well $R_0$ and $\Theta_0$ measurements, are crucial for correcting the Galactic acceleration effects. 
These measurements can hopefully be improved in the future, for example with the astrometric data from \emph{Gaia}, which has already yielded a rather precise acceleration measurement of the SSB \citep{Klioner2021A&A}. 

The bottom panel of Fig.~\ref{fig:Gal} demonstrates the Shklovskii contribution for proper motion of 2.5 mas/yr (blue) and 5.0 mas/yr (orange) at both directions (assuming a 1\% observing error). 
However, some pulsars have very large proper motions, e.g. PSR~J0437$-$4715 has a proper motion $\sim$140~mas/yr \citep{Reardon2024ApJ}. 
While the proper motion of nearby binary pulsars, such as the Double Pulsar, can usually be measured from pulsar timing with good precision using relatively long datasets, the distance of pulsar is difficult to determine and reach the required precision, especially for distant pulsars. 
In the following, I briefly summarise methods to determine the pulsar distance.

\subsection{Astrometry and pulsar distance}
\label{sec:astrometry}
Distance measurement of celestial objects is one of the most challenging tasks in observational astronomy. 
The classical and still effective method of distance measurement for nearby stars is parallax, where the apparent position of the star relative to the background is observed when the Earth moves to different phases of its orbit around the Sun. 
The semi-angle of inclination between two sight-lines to the star is called parallax ($\pi_x$, measured in arcseconds), whose reciprocal results in a distance measurement in parsecs: $d=1/\pi_x$. 

\begin{figure}[t]
    \centering
    \includegraphics[width=0.55\textwidth]{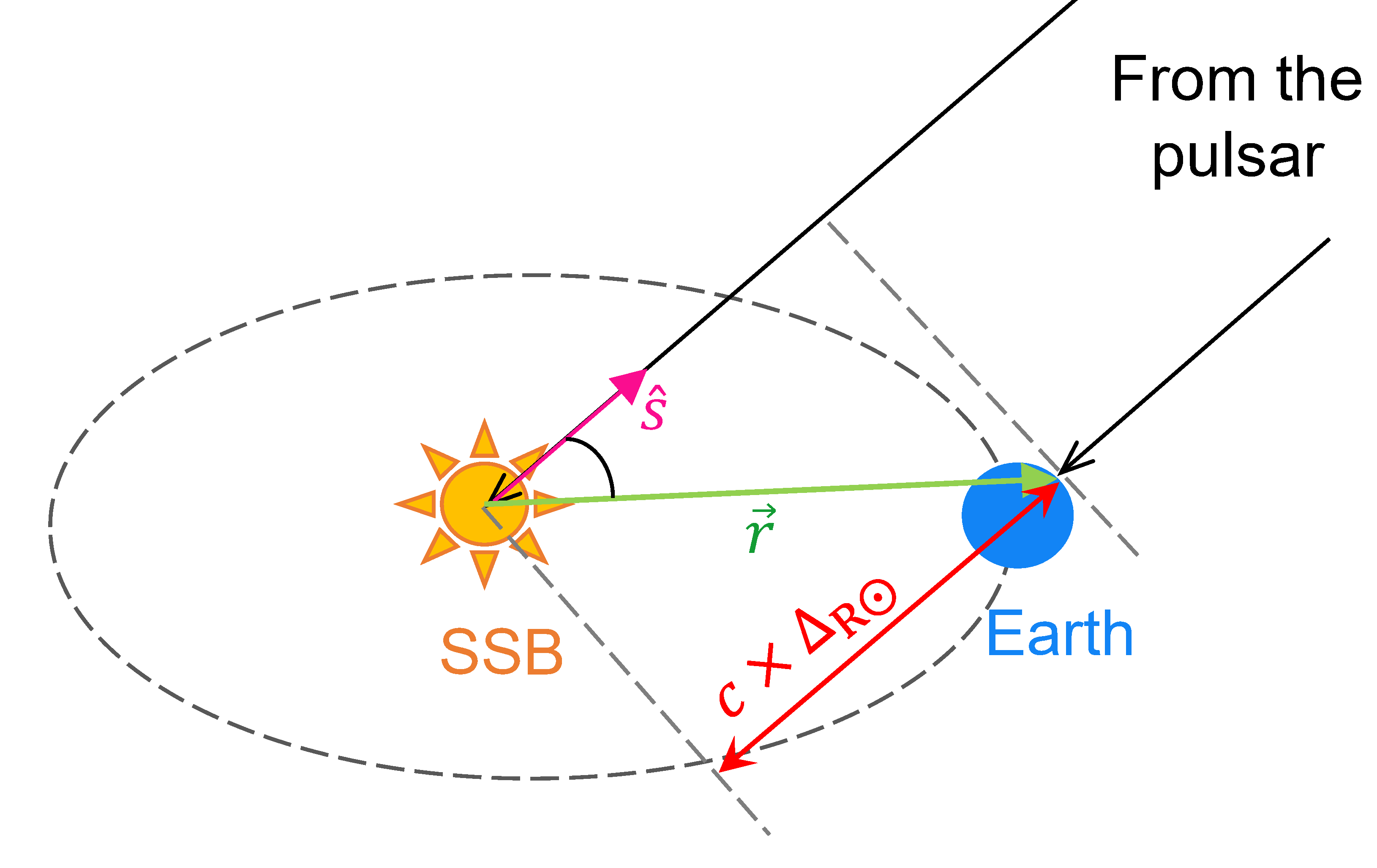}
    \caption{Geometry of R\o{}mer delay between the telescope and the SSB. $\hat{s}$ is a unit vector pointing from the SSB to the pulsar, and $\vec{r}$ is the vector pointing from the SSB to the telescope. The difference of light travel distance from pulsar to the telescope and to the SSB (marked in red) is $c\times \Delta_\mathrm{R\odot} = -\vec{r}\cdot\hat{s} $. Adapted from Fig. 2.7 of \cite{Hu2023PhDT}.}
    \label{fig:tssb}
    \vspace{-5pt}
\end{figure}

Pulsar timing allows not only precise measurements of the rotation of pulsars and their motions (e.g. in a binary orbit) precisely, but also astrometric measurements (position and proper motion in right ascension and declination, and timing parallax) by transferring the topocentric arrival time of pulses observed at radio telescopes throughout the year to the SSB.
The pulsar position is derived from the annual variation of the pulse arrival due to the change of R\o{}mer delay \citep{Romer1676}---the light travel time between the phase centre of the telescope and the SSB: 
\begin{equation}
    \Delta_{\mathrm{R}\odot} = -\frac{1}{c} \, \vec{r}\times\hat{s} \,,
\end{equation}
where $\vec{r}$ is the vector connecting the SSB to the observatory and $\hat{s}$ is a unit vector pointing from the SSB to the position of the pulsar. An illustration is shown in Fig.~\ref{fig:tssb}.
The spatial motion of the pulsar with respect to the SSB manifests as transverse velocity and measured as proper motion ($\mu_\alpha$ and $\mu_\beta$) in the equatorial coordinate system. 
Finally, an annual parallax may be presented in the timing residuals of nearby pulsars with high timing precision \citep{Backer_Hellings_1986ARA&A}:
\begin{equation}
\Delta t_\pi = -\frac{1}{2cd}(\vec{r}\times\hat{s})^2 \,.
\end{equation}
The timing parallax is a measurement of the curvature of the emitted wavefronts at different positions of the Earth’s orbit. 
This effect causes a change in the arrival time of the pulse with an amplitude of $l^2\cos{\beta}/(2cd)$, where $l$ is the Earth-Sun distance and $\beta$ is the ecliptic latitude of the pulsar. 
This means that timing parallax can only be measured for pulsars close to the ecliptic plane, other pulsars will require additional independent approaches such as VLBI \citep{Gwinn1986AJ} or space astrometry such as \emph{Gaia} \citep{Jennings+2018ApJ}, if there is a bright optical companion in the system. 

VLBI offers high-resolution radio imaging due to the large baselines between participating telescopes, enabling precise, model-independent measurements of pulsar distances via annual geometric parallax. 
The most accurate VLBI parallax measurements to date was on PSR~J2222$-$0137, $\pi_x= 3.723^{+0.013}_{-0.014}$~mas, with an accuracy of less than 0.4\% \citep{Ding+2024,Guo+2021,Deller2013ApJ}.
However, VLBI parallax measurement also becomes difficult for distant pulsars. 

As an alternative, the distance of pulsars can be estimated from DM using Galactic electron density models such as NE2001 \citep{NE2001} or YMW16 \citep{YMW2017ApJ}. 
However, these estimates do not provide uncertainty and vary from one to another.
For instance, in the case of the Double Pulsar, the NE2001 model suggests a distance of 516~pc, whereas the estimate from the YMW16 model changes from 1105~pc to 463~pc when the VLBI-based distance of 1100 pc \citep{Deller+2009Sci} is removed from the list of independent distances in the model calibration \citep[see][for details]{Kramer+2021PRX}. 
Both of these DM distances are significantly different from the VLBI measurement of $770\pm 70$~pc, but are consistent with the timing parallax distances \citep[][Hu et al. in prep.]{Kramer+2021PRX}. 
Possibly due to different systematic errors and reference frames, there are discrepancies in the distance of many pulsars between different approaches, even the nearest millisecond pulsar, J0437$-$4715 \citep{Reardon2024ApJ,Jennings+2018ApJ}. 
For many binary pulsars, such as PSR~J1913+1102 and PSR~J1946+2052, the lack of knowledge of pulsar distances is a limiting factor for GW tests. 

\begin{figure}
    \vspace{-10pt}
    \centering
    \includegraphics[width=0.7\linewidth]{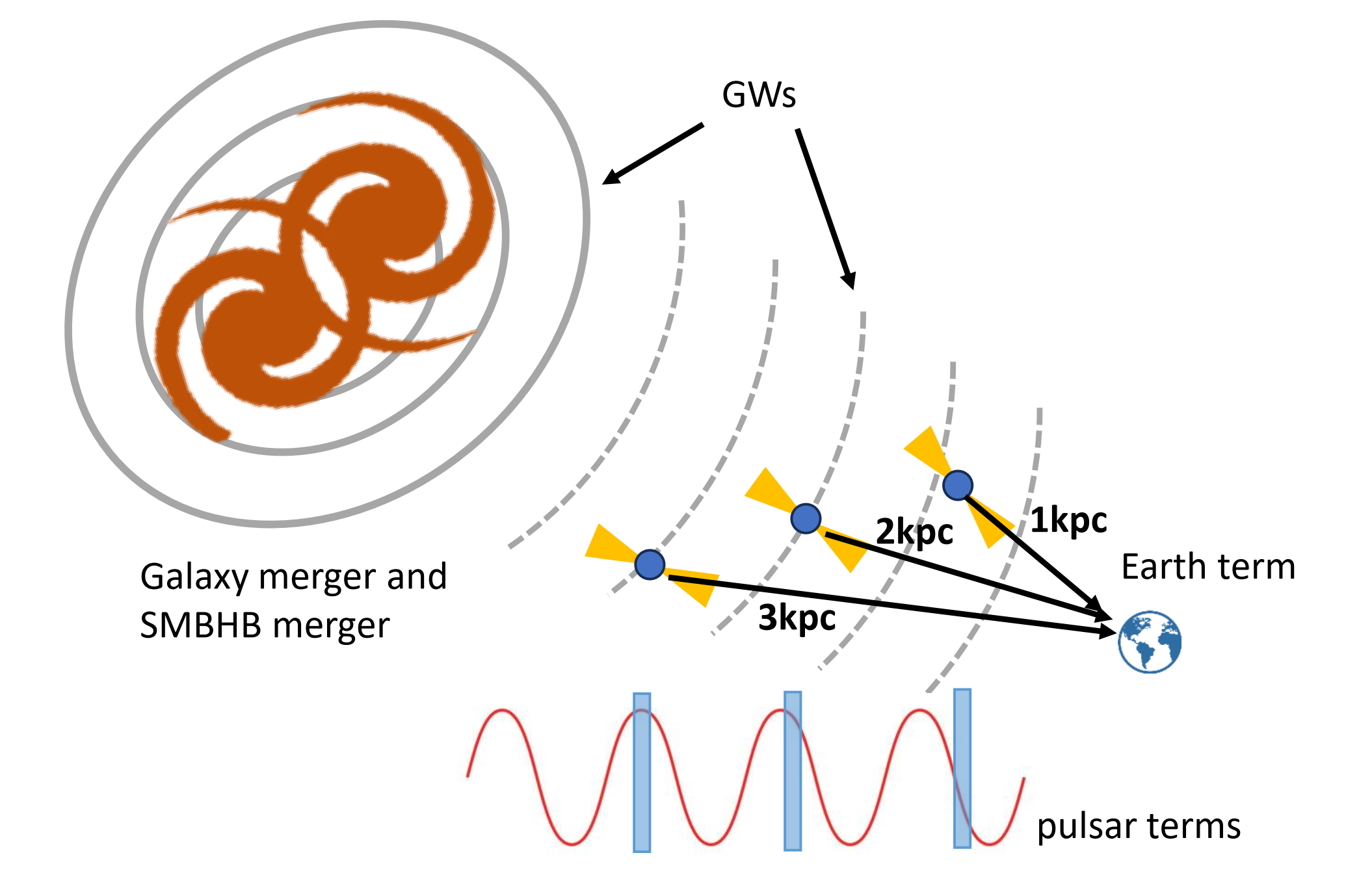}
    \caption{Studying SMBHB evolution over thousands of years using pulsar terms. }
    \label{fig:GWterm}
    \vspace{-10pt}
\end{figure}

Moreover, pulsar distance is also important in the PTA analysis. 
A major challenge in finding the common “red” signal from the GWB is its strong correlation with astrometric parameters (including parallax) and DM modelling from pulsar timing. 
Second, precise knowledge of pulsar distances is essential for the analysis of the pulsar term\footnote{Current GWB studies focus on the Earth term, the effect of the GW at Earth when the pulse is received.}, i.e. the contribution to the timing residuals from the GW signal at the pulsar's location when the pulse was emitted thousands of years ago \citep{Detweiler1979,Jenet2005}. 
Due to different positions of pulsars relative to the Earth and the GW source, the pulsar term probes different evolutionary stages of SMBHBs (see Fig.~\ref{fig:GWterm} for an illustration) and gives access to the evolutionary history and spin of SMBHBs \citep{Mingarelli2012PhRvL}. 
As demonstrated by \cite{Kato2023}, precise distance measurements from VLBI observations can dramatically enhance the accuracy of GW source localization. 
Independent distance measurements, with e.g. VLBI \citep{Ding2023} and \emph{Gaia}, are therefore much favoured in order to break correlation and increase the PTA sensitivity. 

It has been shown in \cite{LiuKuo2025} that an inappropriate solar wind model can introduce significant bias in astrometry parameters, which is highly dependent on the ecliptic latitude of the pulsar and the timing precision of the data. 
In the following, I will demonstrate through simplified simulations how red noise and imperfect DM modelling affect astrometry and timing parameter measurements.

\subsection{Modelling red noise}

\begin{figure}[t]
    \vspace{-10pt}
    \centering
    \includegraphics[width=\textwidth]{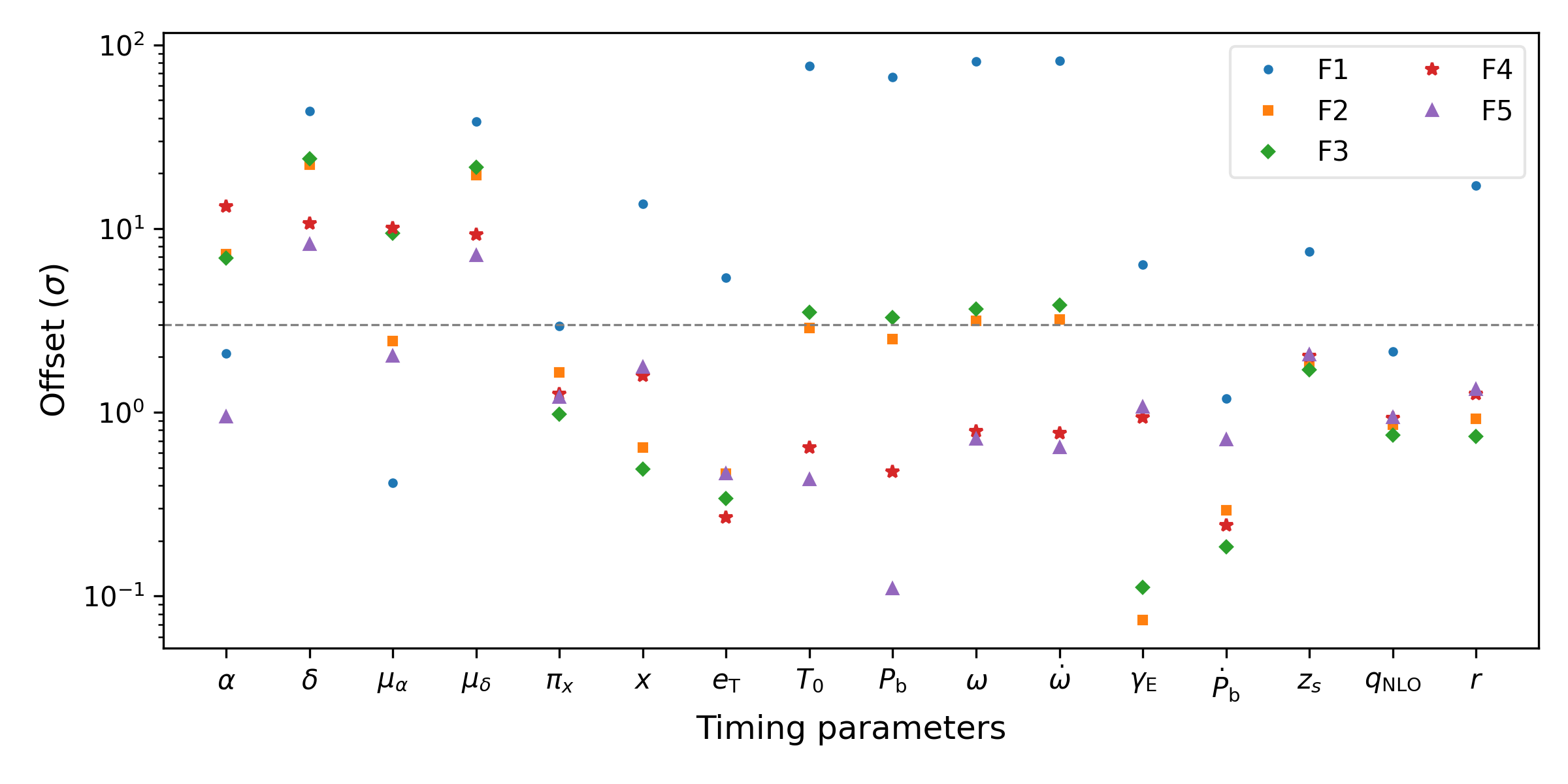}
    \vspace{-15pt}
    \caption{Comparison of timing parameter offsets compared to input values when fitting for different maximum order of spin frequency derivative parameters (in absolute terms). The blue, orange, green, red, and purple points represent F1, F2, F3, F4, and F5, respectively. 
    From left to right are: right ascension $\alpha$, declination $\delta$, proper motion $\mu_\alpha$ and $\mu_\delta$, timing parallax $\pi_x$, projected semi-major axis $x$, orbital eccentricity $e_\mathrm{T}$, epoch of periastron $T_0$, orbital period $P_\mathrm{b}$, longitude of periastron $\omega$, relativistic advance of periastron $\dot{\omega}$, Einstein delay amplitude $\gamma_\mathrm{E}$, orbital period derivative $\dot{P}_\mathrm{b}$, 
    logarithmic Shapiro shape parameter $z_s=-\ln(1-s)$, NLO factor for signal propagation $q_\mathrm{NLO}$, and Shapiro range parameter $r$.
    The grey dashed line indicates $3\sigma$.}
    \label{fig:Fn}
    \vspace{-10pt}
\end{figure}

In this simulation, I inject a red noise with an amplitude of $A=1\times 10^{-12}$ and a spectral index of $\gamma=2$, following the power spectral density formula $P(f) = A^2 f^{-\gamma}/ 12 \pi^2 $, in a 5-yr long dataset of PSR~J0737$-$3039A and assume no variation in DM. 
The fitting results of astrometric and binary parameters using different maximum order of spin frequency derivative parameters (F1-F5) are shown in Fig.~\ref{fig:Fn}. 
The vertical axis shows offsets of fitted parameters compared to the input values in the simulation. 
To make the results statistically significant, 1000 simulations were performed. 

By fitting to different maximum Fn parameters, the offset of timing parameters could vary by two orders of magnitude, indicating a strong correlation with red noise modelling. 
In general, the offset of most parameters to its input value reduces when more spin frequency derivatives are used. 
Binary parameters are well recovered after fitting for F4, with offsets much below 2$\sigma$.  
Timing parallax recovers to just 1$\sigma$ away from the input value. 
The only parameters not recovered are the declination $\delta$ and proper motion $\mu_\delta$, which still differ by $>7\sigma$ after fitting for F5. 
This is probably because the pulsar is not exactly in the ecliptic plane.

\subsection{Modelling DM variation}
\label{sec:dmvar}

\begin{figure}[t]
    \vspace{-10pt}
    \centering
    \includegraphics[width=\textwidth]{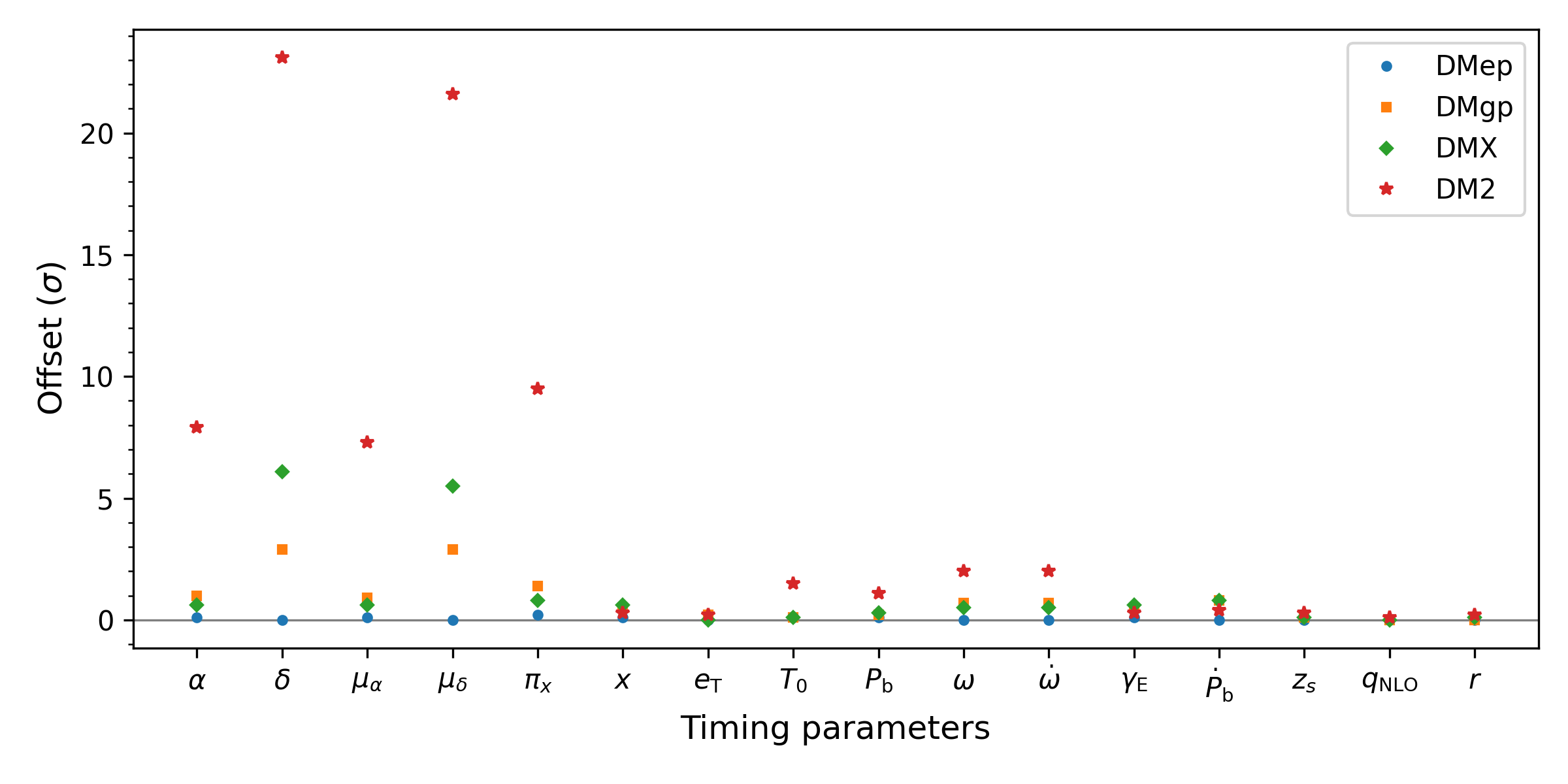}
    \vspace{-15pt}
    \caption{Offset of timing parameters using different DM models. The blue points use DM correction based on single epoch DM measurements (DMep), the orange squares use a Gaussian process model (DMgp) based on DMep, the green diamonds use a DMX model (30 days), and the red star fit for DM1+DM2.}
    \label{fig:DMsim}
    \vspace{-10pt}
\end{figure}

Here I perform similar simulations but assuming no red noise and taking the DM variation from real MeerKAT observations \citep{Hu+2022}. 
These DM values were measured per observing epoch while keeping the rest of parameters to the best known values \citep{Kramer+2021PRX,Hu+2022}. 
The offset of this DM variation to a reference DM is written in a text file, which is then read by a modified \textsc{tempo} version developed in \cite{Kramer+2021PRX} to correct DM on a per epoch basis. 
As DM measurements from MeerKAT UHF and L bands are very precise, per epoch DM correction (DMep) should in principle have minimal affect on timing parameters compared to classical DM modelling methods that suffer from correlation. 

Figure~\ref{fig:DMsim} shows the offset of fitted timing parameters compared to their input values when different DM modelling was employed: DMep (blue), Gaussian process model (DMgp) based on DMep (orange), DMX model with the same cadence as observations (green), and DM derivative model with fitting for DM1+DM2 (red). 
Again, 1000 Monte Carlo simulations were performed to sample the uncertainties in DM. 

The figure shows that indeed DMep performs the best on all parameters with the minimum offset compared to input values, even the timing parallax is well recovered. 
DMgp performs a bit worse, especially on $\delta$ and $\mu_\delta$, but all parameters are still within $3\sigma$ difference to the input values. 
In contrast, the DMX and DM derivative models yield less accurate values as they are fitted simultaneously with timing parameters and are especially strongly correlated with astrometry parameters. 
While the DMX model is still able to recover most of parameters (except for $\delta$ and $\mu_\delta$), the DM2 model is deficient in recovering astrometric parameters including parallax and performs worse on some of binary parameters.

Based on these simulations, one finds that compared to red noise, in general DM modelling has less impact on timing parameters and can be minimised with the DMep model.

\section{Concluding remarks}\label{conclude}

Recent developments in pulsar-based experiments have continued to affirm the unique role that pulsars---especially those in relativistic binaries---play in testing fundamental physics. 
The ongoing precision timing of systems like the Double Pulsar not only provides some of the most stringent tests of GR in the strong-field regime but also opens new avenues for probing the EOS of dense matter. 
Meanwhile, PTAs are entering an era of discovery, offering direct insights into the nanohertz GW universe. 
Despite these advances, several challenges remain, including the need for improved distance measurements, accurate and reliable Galactic potential model, and refined DM and noise modelling. 
They are crucial for exploring the boundary of GR, providing robust measurements of neutron star MOI, as well as identifying the sources of the GWB and potential single bright sources. 

Additional challenges include, but not limited to, data combination from distinct datasets, clock errors leading to monopolar signals in the GW analysis, timing model inaccuracies (e.g. due to unmodelled binary motion or glitches), solar wind variability \citep{Tiburzi2021A&A}, and errors in the Solar System ephemerides \citep{Vallisneri2020}. 
However, a comprehensive review is beyond the scope of this article.

It is also worth noting that a recent work \citep{Hu+2025AA} has pointed out that inappropriate data processing can lead to smearing of the pulse profiles, resulting in apparent orbital DM variations. 
Such effects are prominent at low frequencies (e.g. with SKA-Low) and should be treated with care.
In addition, as modern telescopes are more sensitive, the accuracy of current pulse phase predictors (used for folding) starts to be limited and may bias the Shapiro delay parameters. 
Advanced pulsar data processing software are desired in the coming years as large telescopes and arrays are built.

Furthermore, in addition to the FAST Core Array, several radio telescopes in excess of the 100-metre class are under construction in China and expected to be completed by 2028: 
the 110-m Qitai radio Telescope (QTT) in the north-west \citep{QTT2023}, Jingdong 120-m pulsar Radio Telescope \citep[JRT,][]{JRT2022} in the south-west, and another 120-m telescope located in Huadian (Jilin Province) in the north-east. 
The pathfinding pulsar observations with the Chinese VLBI Network (CVN) incorporating the FAST have demonstrated the potential of astrometric measurements of the CVN \citep{YanZhen2024ChPhL}. 
Integrating these giant telescopes into the CVN will not only improve observational accuracy, but also improve the angular resolution of the imaging, given that their longest baselines are over 3000~km.

The integration of the SKA with VLBI networks, commonly referred to as SKA-VLBI, will mark a major leap in astrometric precision and sensitivity. 
By phasing up SKA with global VLBI arrays, the sensitivity will be improved by an order of magnitude, push astrometric precision to $\mu$as-level for a much larger number of radio sources, including millisecond pulsars \citep{LiYingjie2024RAA}. 
This will dramatically expand the catalogue of pulsars with model-independent distance and proper motion measurements, a crucial input for high-precision pulsar timing and GW detection. 

In the era of new-generation radio telescopes, continued pulsar observations and surveys over the next few years are set to transform gravity and GW research using radio pulsars, with the potential to reveal new physics beyond GR.

\backmatter

\bmhead{Acknowledgements}
I am grateful to the anonymous referee for carefully reading the manuscript and providing helpful suggestions, which improved the quality of this work.
I would like to thank Michael Kramer, Norbert Wex, David Champion, Xueli Miao, and Junjie Zhao for valuable discussions. 
I owe special thanks to Norbert Wex for providing several figures that improved the visualisation of this paper. 
I deeply appreciate the Matter and Cosmos Section (SMuK) of the German Physical Society (DPG), and to those that nominate the awardee and serve on the committee, for their recognition of her dissertation work.

\section*{Declarations}

\begin{itemize}
\item Funding: Supported by the Max Planck Society.
\item Competing interests: The author declares no competing interests.
\item Author contribution: Sole contributor to the preparation and writing of this work.
\end{itemize}

\noindent

\bibliography{mybib}

\end{document}